\documentclass[useAMS,usenatbib,a4paper]{mn2e}

\usepackage{natbib}
\usepackage{graphicx}
\usepackage{color}
\usepackage{amssymb}
\usepackage{amsmath}
\usepackage{xspace}
\usepackage{xargs}
\usepackage{ulem}
\usepackage{verbatim}
\usepackage[usenames,dvipsnames]{xcolor}
\usepackage{subfigure}

\definecolor{FireBrick}{RGB}{178, 34, 34}

\newcommand{\rk}[1]{{\textcolor{black}{#1}}}

\newcommand{\myemail}{Rebecca.Kennedy@nottingham.ac.uk}
\newcommand{\galapagos}{{\scshape galapagos}\xspace}

\newcommand{\galfitthree}{{\scshape galfit3}\xspace}
\newcommand{\galfitm}{{\scshape galfitm}\xspace}
\newcommand{\megamorph}{{MegaMorph}\xspace}
\newcommand{\sersic}{S\'ersic\xspace}

\newcommandx{\N}[2][1= ,2= ]{$\mathcal{N}^{#1}_{#2}$\xspace} 
\newcommandx{\R}[2][1= ,2= ]{$\mathcal{R}^{#1}_{#2}$\xspace}
\newcommand{\re}{R_{\rm e}}
\newcommandx{\Rss}{$\mathcal{R}_{\rm ss}$\xspace}
\newcommandx{\Nss}{$\mathcal{N}_{\rm ss}$\xspace}
\begin{document}
\title[Galaxy structure with bulge-disc decompositions]{Galaxy And Mass Assembly (GAMA): Understanding the wavelength dependence of galaxy structure with bulge-disc decompositions}
\author[Kennedy et al.]{Rebecca~Kennedy$^{1}$\thanks{E-mail: \myemail}, Steven~P.~Bamford,$^{1}$ Boris~H\"au{\ss}ler$^{2,3,4}$, Ivan Baldry$^{5}$,\newauthor Malcolm~Bremer$^{6}$, Sarah~Brough$^{7}$, Michael~J.~I.~Brown$^{8}$, Simon~Driver$^{9,10}$,\newauthor Kenneth~Duncan$^{1,11}$, Alister~W.~Graham$^{12}$, Benne~W.~Holwerda$^{11}$,\newauthor Andrew~M.~Hopkins$^{7}$, Lee~S.~Kelvin$^{5}$, Rebecca~Lange$^{9}$, Steven~Phillipps$^{6}$,\newauthor Marina~Vika$^{13}$ and Benedetta~Vulcani$^{14}$
\smallskip\\
$^{1}$School of Physics \& Astronomy, The University of Nottingham, University Park, Nottingham, NG7 2RD, UK\\
$^{2}$University of Oxford, Denys Wilkinson Building, Keble Road, Oxford, Oxon OX1 3RH, UK\\
$^{3}$University of Hertfordshire, Hatfield, Hertfordshire AL10 9AB, UK\\
$^{4}$European Southern Observatory, Alonso de Cordova 3107, Vitacura, Casilla 19001, Santiago, Chile\\
$^{5}$ARI, Liverpool John Moores University, IC2, Liverpool Science Park, 146 Brownlow Hill, Liverpool, L3 5RF\\
$^{6}$School of Physics, HH Wills Physics Laboratory, Tyndall Avenue, Bristol BS8 1TL, UK\\
$^{7}$Australian Astronomical Observatory, PO Box 915, North Ryde, NSW 1670, Australia \\
$^{8}$School of Physics and Astronomy, Monash University, Clayton, Victoria 3800, Australia\\
$^{9}$ICRAR, The University of Western Australia, 35 Stirling Highway, Crawley, WA 6009, Australia\\
$^{10}$SUPA, School of Physics and Astronomy, University of St. Andrews, North Haugh, St. Andrews KY16 9SS, UK\\
$^{11}$Leiden Observatory, Leiden University, NL-2300 RA Leiden, Netherlands\\
$^{12}$Centre for Astrophysics and Supercomputing, Swinburne University of Technology, Victoria 3122, Australia\\
$^{13}$IAASARS, National Observatory of Athens, GR-15236 Penteli, Greece\\
$^{14}$Kavli Institute for the Physics and Mathematics of the Universe (WPI), UTIAS, the University of Tokyo, Kashiwa, 277-8582, Japan\\
}
\maketitle
\begin{abstract}

With a large sample of bright, low-redshift galaxies with optical$-$near-IR imaging from the GAMA survey we use bulge-disc decompositions to understand the wavelength-dependent behavior of single-\sersic structural measurements.

We denote the variation in single-\sersic index with wavelength as \N, likewise for effective radius we use \R. We find that most galaxies with a substantial disc, even those with no discernable bulge, display a high value of \N. The increase in \sersic index to longer wavelengths is therefore intrinsic to discs, apparently resulting from radial variations in stellar population and/or dust reddening. Similarly, low values of \R ($<$ 1) are found to be ubiquitous, implying an element of universality in galaxy colour gradients.

We also study how bulge and disc colour distributions vary with galaxy type. We find that, rather than all bulges being red and all discs being blue in absolute terms, both components become redder for galaxies with redder total colours. We even observe that bulges in bluer galaxies are typically bluer than discs in red galaxies, and that bulges and discs are closer in colour for fainter galaxies. Trends in total colour are therefore not solely due to the colour or flux dominance of the bulge or disc.

\end{abstract}


\section{Introduction}

The formation history of a galaxy is recorded in the age, metallicity and phase space distribution of its stellar populations. The stars in a given galaxy have formed over a range of times and through different mechanisms, so the observed spatial structure of a galaxy, and its wavelength dependence, can be used to learn how galaxies formed and evolved.

Ideally we would be able to decompose a galaxy into all its constituent components, but this is currently not possible. Instead, we are able to do bulge-disc decompositions that at least allow us to differentiate between central and extended components within a bright sample.  In the case of distant or faint galaxies with low signal-to-noise we may only be able to fit a single component, but this still has its merits if we can ensure that our interpretation of single-component fits is consistent with bulge-disc decompositions \citep{Allen2006}.

\rk{Historically, galaxies were visually classified as elliptical (`E'), lenticular (`S0') or spiral (`S') in morphology \citep{Hubble1936}. Elliptical galaxies are traditionally characterised by their one-component spheroidal shape and collapsed structure. They are thought to be the product of early and/or dry mergers, and therefore contain older, redder stars \citep{Dressler1997,Kauffmann2003,Brinchmann2004}. Many of the objects initially classified as `elliptical' were later found to have a disc component, which led to the introduction of the `ES' classification for galaxies lying between ellipticals and lenticulars on the Hubble tuning fork diagram \citep{Liller1966}. They have since been referred to as E/S0 galaxies and discy ellipticals \citep{Nieto1988,Simien1990S}. This view was later augmented to include parallel sequences for spirals and lenticulars, with `early' and `late' types described by their disc-to-bulge ratios \citep{VandenBergh1976}. As the quality of observational data improved over time, the need for a continuum of bulge-to-disc ratios in early-type galaxies also became necessary \citep{Capaccioli1988}. This version of the Hubble diagram was then extended to include spheroidal galaxies at the end of the S0a-S0b-S0c sequence \citep{Cappellari2011}.}
\rk{Morphological classifications based on kinematics are more sensitive to the presence of discs than any photometric attempt \citep[e.g.]{Emsellem2007,Cappellari2011}. Kinematics can also distinguish between, for example, fast and slow rotating early-type galaxies, which can shine a light on both their underlying stellar structure, and their possible formation mechanisms \citep[e.g.][and references therein]{Emsellem2011}.}

\rk{An alternative classification system is proposed in \cite{Graham2014}, and expanded upon in Graham et al. (submitted), in which both bulge-to-disc flux ratio and Hubble type are used in conjunction with one another, in order to minimise the effect of the random orientation orientation of a galaxy's disc on its morphological classification.}

Radial luminosity profiles of galaxies, and their components, are commonly described by a \sersic index, $n$, which models the variation in the projected light distribution with radius \citep{Sersic1963,Graham2005}.

However, using a single-\sersic fit as an indicator of whether a galaxy is early- or late-type can sometimes be misleading, due to the intrinsic variations in $n$ and measurement uncertainties \citep{Binggeli1998, Graham2003, Krajnovic2013}.
Elliptical galaxies commonly have discs \citep[e.g.][and references therein]{Kormendy1996}, or disc-like structures \citep{Emsellem2007,Krajnovic2011,Emsellem2011,Cappellari2011}, and can exhibit a large range of disc-to-total (D/T) flux ratios \citep{Krajnovic2013}, with $D/T \sim 0.4$ typical.  This calls into question the tradition of classifying galaxies by their \sersic index or morphology \citep{Vika2015}.

Similarly, a number of galaxies that have previously been classified as spirals show no morphological evidence of a classical bulge, and instead have an irregular central bright component \citep{Carollo1997}.

Discs can generally be described by an `exponential' profile, with $n \sim 1$, whilst classical bulges and elliptical galaxies are generally described by a higher \sersic index of $n \gtrsim 2.5$ (e.g. \citealp{Graham2013}).  In fact, it is rare to find bulges with $n > 3$ \citep{Balcells2003}, and the bulges of many spiral galaxies (particularly intermediate-type discs) exhibit the exponential profiles of pseudo bulges and are supported by rotation \citep{Andredakis1994, Carollo1999, deJong2004, Gadotti2009, McDonald2011}.  They generally have younger stellar populations than classical bulges, and are likely formed by secular processes.  Due to their flattened light profiles they are often difficult to detect at high inclination \citep{Carollo1997,Kormendy2006,Drory2007,Gadotti2009}. Conversely, early-type spiral galaxies appear to have significantly bigger and brighter bulges than late-type spiral galaxies, which tend to have small, faint bulges \citep{Graham2001, Mollenhoff2004}. A possible reason for this difference in bulge size with morphology is that in late-type spirals the bulge is `submerged' in the disc, masking some of the bulge light \citep{Graham2001}. This is consistent with the Hubble sequence pattern of increasing D/T flux ratio for later-type spirals \citep{DeLapparent2011}.

Previous studies have shown a strong relationship between measured sizes of galaxies and wavelength; on average, galaxies of all morphologies are found to be smaller in redder wavebands (\citealt{Evans1994,LaBarbera2010b,McDonald2011,Kelvin2012}; \citealt{Vulcani2014} (hereafter V14); \citealt{Kennedy2015}). 
\sersic index is also known to change with wavelength; \sersic index measured in the NIR is generally significantly larger than in the optical for low-$n$ galaxies, with high-$n$ galaxies exhibiting a similar, if less pronounced, trend (\citealt{Taylor-Mager2007,Kelvin2012}; V14).

The colour of a galaxy and its components can be used to differentiate between early and late morphological types, at $u-r = 2.22$, regardless of magnitude \citep{Strateva2001}. $u-r$ versus $g-i$ colour space has also been found to effectively classify galaxies in the Sloan Digital Sky Survey (SDSS) as either early- or late-type. The location of a galaxy in this colour space also reflects the degree and locality of star formation activity \citep{Park2005}, and correlates well with stellar population age.

Colour gradients within galaxy components may also prove to be an effective classifier; star-forming galaxies do not show significant colour gradients in their discs, whilst passive galaxies do.  These gradients may be due to dust extinction; the discs of active galaxies are optically thin, resulting in no colour gradient, whilst galaxies that \textit{do} show a colour gradient appear optically thick in the centre and optically thin in their outer regions \citep{Cunow2001}.  Meanwhile, ellipticals are thought to form with steep stellar population gradients (see e.g. \citealt{Brough2007,Kuntschner2010} for more discussion).

V14 examined the variation of \sersic index ($n$) and effective radius ($\re$) with wavelength in order to reveal the internal structure, and therefore the formation history, of galaxies in their sample. As in V14, the notation $\mathcal{N}^{H}_{g} = n(H)/n(g)$ and $\mathcal{R}^{H}_{g} = \re(H)/\re(g)$ will be used here to denote the ratio between the $H$- and $g$-bands. We omit the waveband labels from \N and \R when discussing their general behavior.  V14 speculates that the variation in \N reflects whether a system has one or two components; in a high-\N system we are observing the \sersic index of a disc in bluer wavelengths and a bulge in redder wavelengths.  Conversely, V14 suggests that for one-component systems we see \N closer to unity because we are measuring the \sersic index of just one component at all wavelengths, e.g. in the case of elliptical galaxies.  There is, however, a large change in $\re$ with wavelength for high-$n$ galaxies, which shows that elliptical galaxies contain a radial progression of different stellar populations, possibly resulting from multiple minor merging events throughout the galaxy's lifetime.

\citet{Vika2015} found that by combining \N with the colour information of the galaxy we can separate elliptical galaxies from S0s more reliably than other photometric classification methods.

V14 and \citet{Vika2015} suggest that inferences about a galaxy's bulge-disc nature can be made from single-\sersic fits. In this paper we use a large sample of low-redshift galaxies to study whether there is a connection between bulge-disc properties and single-\sersic results. Using our multi-wavelength bulge$-$disc decompositions we also study the relationship between bulge and disc properties in order to uncover information about the developmental histories of these galaxies.  As in \citet{Kennedy2015} we first ensure that the recovered properties of our bulges and discs are robust with redshift (Section \ref{subsec:z_BD}), before studying the wavelength dependence of $n$ and $\re$ as a function of bulge:total (B/T) flux ratio, in order to determine whether V14's inferences from single-\sersic fits are consistent with bulge-disc decompositions (Section \ref{subsec:wave_dep}).  We go on to look at the relative colours of bulges and discs for six subsamples (as defined in V14), and what they (and their single-\sersic colours) can tell us about their likely formation histories (Section \ref{subsec:B_D_colour}). We then explore the relative colours of these components in the context of visual morphological type (Section \ref{subsec:B_D_morph}),  before examining trends in physical properties as a function of luminosity (Section \ref{subsec:B_D_lum}).

The analysis has been carried out using a cosmology with ($\Omega_{m},\Omega_{\Lambda}, h) = (0.3, 0.7, 0.7)$ and AB magnitudes.

\section{Data}

The sample of galaxies used in this project, and their single-\sersic structural measurements, have previously been presented in \citet{Haussler2013} and studied further in V14 and \citet{Kennedy2015}. A detailed description of the selection criteria, robustness of fits and properties of the sample can be found in those papers; here, a brief overview is given.

The sample used in this project is taken from the G09 region of the Galaxy And Mass Assembly (GAMA) survey II \citep{Driver2009,Driver2011,Liske2015}, which is the largest homogeneous multi-wavelength dataset currently available. GAMA includes data from both SDSS (\citealt{York2000}) and UKIDSS (\citealt{Lawrence2007}), which provide a consistent and complete set of imaging covering the (\textit{ugriz}) optical bands and the near-IR (\textit{YJHK}) bands. It has been demonstrated that, for our sample limits, all these bands have a depth and resolution that allows for \sersic-profile fitting (\citealt{Kelvin2012}).

\subsection{K-correction and rest-frame cuts}
\label{subsec:K_corr_etc}
To obtain rest-frame colours for the bulge and disc components of both samples, K-corrections have been performed using the SED fitting code of \citet{Duncan2014}. Following a method similar to that of \citet{Blanton2007}, stellar population synthesis models from \citet{Bruzual2003} are fit to the decomposed bulge and disc photometry and the rest-frame colours taken from the best-fit model for each component. The model stellar populations are drawn from a wide range of ages, star-formation histories and metallicities, with dust attenuation allowed to vary in the range $0 \leq A_{V} \leq 4 $ assuming the \citet{Calzetti2000} attenuation law.  We then apply the following criteria to select reliable fits.

For bulges only:
\begin{itemize}
	\item $0.201 < n_{B} < 7.75$, to eliminate values that lie very close to the fitting boundaries.
\end{itemize}

For all galaxies (largely the same as those used in H13 and V14):
\begin{itemize}
	\item $0 < m_{B,D} < 40$ at all wavelengths, where $m_{B,D}$ is the total apparent magnitude in each band for the bulge and the disc, respectively.
    \item $m - 5 < m_{B,D} < m + 5$, where m is the starting value of the magnitude in each band. See \cite{Haussler2013} for more details.
    \item $0.301 < R_{e}(B,D) < 399$ pixels, which ensures bulge and disc sizes remain in a physically meaningful range.
    \item $0.001 < q_{B,D} \leq 1.0$, where $q_{B,D}$ is the axial ratio of the bulge and disc, respectively.
    \item Position ($x$, $y$); positions are constrained to lie within a box of size $0.5R_{\rmn{e}}$ around the centre as defined by the single-\sersic fit.  Additionally, the position of the disc and the bulge are constrained to be the same.
\end{itemize}

\begin{figure}
		\centering
		\includegraphics[width=0.4\textwidth]{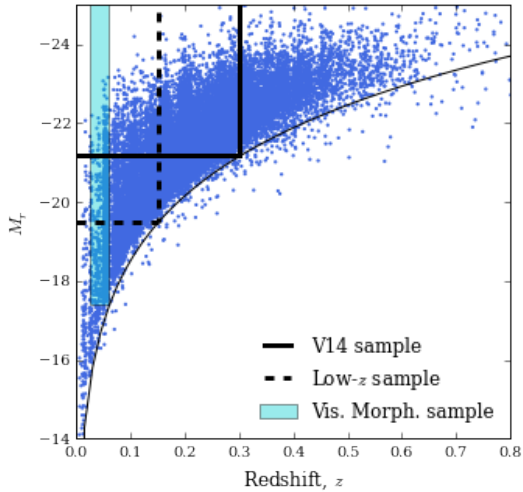}		
		\caption{Absolute $r$-band magnitude versus redshift for our parent sample, with our volume-limited selection boxes overlaid.  The thinnest black line indicates the primary apparent magnitude limit of the GAMA II redshift survey, $r < 19.8$. This corresponds to an absolute magnitude of $M_{r} = -21.2$ at $z = 0.3$ and $M_{r} = -19.48$ at $z = 0.15$. See Table \ref{table:Cleaning1} for the number of galaxies in each sample.
        \label{fig:Sample_selection}}
	\end{figure}

In this study we use two different volume-limited samples (see Fig. \ref{fig:Sample_selection}), which are determined by the the apparent magnitude limit of the GAMA II redshift survey, $r < 19.8$ mag, at two different redshifts. When studying variation in galaxy properties with redshift, a volume-limited sample is taken with $z < 0.3$, $M_{r} < -21.2$ mag, in line with V14. When studying variation in galaxy properties with absolute magnitude, $M_{r}$, a $z < 0.15$, $M_{r} < -19.48$ mag volume-limited sample is taken, allowing galaxies to be observed over a wider range of absolute magnitudes, at the cost of a smaller redshift range. We also use a sample of morphologically classified galaxies from \cite{Kelvin2014} with $M_{r} < -17.4$ mag and $0.025 < z < 0.06$. The number of objects, and `strong' bulge and disc components \rk{(See section~\ref{subsec:Comp_selection} for our definition of `strong')}, in each sample are given in Table \ref{table:Cleaning1}.

Rest-frame colour cuts are made using the single-\sersic magnitudes of each galaxy, in concordance with Figure 3 of V14 and Figure 2 of \cite{Kennedy2015}: we separate our sample into `red' and `blue' galaxies at $u-r = 2.1$, and then subdivide the `blue' sample at $u-r = 1.6$ into `green' and `blue' in order to separate the bluest, potentially starburst, population.  Note that our `green' sample corresponds to the main population of star-forming galaxies, not the green valley.  We also split our sample by \sersic index at $n_{r} = 2.5$, in an effort to separate discy galaxies from ellipticals.

\begin{table}
\centering
	\begin{tabular}{| p{1.7cm} | p{0.9cm} | p{0.9cm} | p{0.9cm} | p{2.2cm} |}
		\hline
		& Pre-cleaning & \textit{Strong} bulges & \textit{Strong} discs & \textit{Strong} bulge \& \textit{strong} discs \\ \hline
		V14   		& 10491 & 5459 & 4456 & 1836 \\
		Low-$z$ 	& 4109 & 2342 & 1945 & 966 \\
		Vis. Morph. & 1013 & 634 & 472 & 264 \\
	\end{tabular}
	\caption{Table showing the number of galaxies in each volume-limited sample. V14 sample: $z < 0.3$, $M_{r} < -21.2$ mag; Low-$z$ sample: $z < 0.15$, $M_{r} < -19.48$ mag; Vis. Morphology sample: $0.025 < z < 0.06$, $M_{r} < -17.4$ mag. Cleaning is applied in all bands simultaneously. The \textit{strong} bulge category contains only bulges which are no more than 3 magnitudes fainter than their corresponding disc (and vice versa for the \textit{strong} disc category). `\textit{Strong} bulge \& \textit{strong} disc' contains galaxies which have both a bulge and disc within three magnitudes of one another, and are therefore a subset of the previous two categories.
		\label{table:Cleaning1}}
\end{table}

\subsection{Structural models}
\label{subsec:Struct_models}
We utilise \galapagos-2 to obtain our structural measurements, which in turn makes use of \galfitm, a version of \galfitthree \citep{Peng2002,Peng:2009fu} extended by the \megamorph project to fit a single wavelength-dependent model simultaneously to many images of a galaxy (\citealt{Haussler2013}; \citealt{Vika2013}; Bamford et al., in prep.). Fitting a light profile in multiple wavebands at once has been found to increase the accuracy and stability of each measurement; \galfitm fits the coefficients of a polynomial function, and the order to which the polynomial can vary can be set by the user. In our single-\sersic model fits, the galaxy magnitudes are allowed to vary freely, whereas \sersic index and effective radius are constrained to be second-order polynomials of wavelength (see \citealt{Haussler2013} and V14 for details).

In addition to the single-\sersic fits presented previously, we have also performed two-component fits, comprising independent \sersic and exponential (a \sersic profile with $n=1$) components (H\"au{\ss}ler et al., in prep), the \sersic index and effective radius of which is constant with wavelength.  We have performed various tests to ensure that our fits consistently move away from their starting parameters, and converge on final solutions that are generally independent of these initial values. The robustness of our decompositions will be discussed in depth in H\"au{\ss}ler et al. (in prep.).  The two components are intended to model the bulge and disc structures seen in many galaxies, and we will often use these labels for convenience, although the interpretation of the two components may vary for galaxies that do not correspond to this simple structural approximation.

\rk{We acknowledge that there may not be statistical evidence for choosing a 2-component fit over a single-\sersic fit. However, one of the problems involved with choosing a 1- or 2-component model is that this builds a dichotomy into the data. One of the strengths of fitting every object with 2 components is that we give consistent treatment to our whole sample, and don't introduce a bias of deciding which fit is more appropriate on a case-by-case basis. Nowadays we know that the vast majority of galaxies are multi-component. Forcing them to be fit by a single component models introduces a bias.  The distributions of component properties we find in this paper strongly support the assumption of multiple components.  This is particularly true for the red galaxies; with single-band fits these can often be well-fit by single-component models, but our multi-band fits clearly indicate a preference for two components with different colours.}

\subsection{Component selection}
\label{subsec:Comp_selection}
In this paper we take a liberal attitude to what constitutes a `bulge', not least because the central component of many of our galaxies is not well resolved.  Thus, bars, lenses, pseudo bulges, classical bulges, and their superpositions, are all swept up in this term.  Our primary aim is to distinguish the extended, thin disk from more centrally concentrated stellar structures.  We postulate that the relative properties of these two components are responsible for much of the observed variation in galaxy properties, particularly that correlated with environment.  We aim to test this claim in detail in future works.

Two-component models have been fit to all galaxies in our sample, regardless of whether they are best modeled as one- or two-component systems.  This raises the issue that one of the components may be negligible, in respect of the luminosity or structure of a galaxy. For example, a small fraction of the light from a one-component elliptical galaxy may be attributed to a disc with poorly-constrained properties, without affecting the resulting model image. A further issue is the potential for one component, or both, to be used to fit some features of a galaxy that they are not intended to model; a false disc may help reduce the residuals caused by an isophotal twist in a pure elliptical galaxy, or a false bulge or a bar may attempt to fit to the arms in a spiral galaxy. To avoid considering the properties of insignificant or incorrect components, a cleaning process can be applied.  Several cleaning methods are employed in the literature, including using a logical filter (e.g. \citealt{Allen2006}), visual inspection (e.g., \citealt{Kelvin2012}) and likelihood-ratio tests (e.g., \citealt{Simard2011}).  While useful, each of these approaches have their difficulties: visual inspection is subjective and insensitive in certain circumstances, whereas goodness-of-fit tests are often unable to eliminate physically meaningless fits.

In our present work we take an extremely simple approach and consider the distributions of component properties at face-value.  We make no special attempt to remove the objects for which a two-component fit is inappropriate, nor do we substitute single-component measurements in any case.  However, we do clean our catalogue of galaxies that may be affected by the constraints imposed on the fit, and hence for which \sersic profile measurements are unlikely to be meaningful.  These criteria are similar to those used in \citet{Kennedy2015}.

We do not consider the poorly-constrained properties of components that make a negligible contribution to the luminosity of their galaxy.  From an examination of the fitting results, we choose to ignore components that are more than three magnitudes fainter than their counterpart (i.e. bulges must have at least 6\% of the luminosity of the corresponding disc to be considered a trustworthy, and therefore `\textit{strong}' bulge, and vice versa), as in \citet{Vika2014}. \rk{See Table~\ref{table:Cleaning1} for the number of galaxies deemed to have a \textit{strong} bulge, \textit{strong} disc, or both a \textit{strong} bulge and \textit{strong} disc.}

\rk{When we use the term `bulges' throughout the paper we are referring to all bulges which are no more than 3 magnitudes fainter than their corresponding disc (i.e. this includes lone bulges/ellipticals AND the bulge components of 2-component galaxies). When we use the term `only \textit{strong} bulges' we are referring to bulges which do not have a significant disc. We similarly use the terms `disc' and `only \textit{strong} disc' throughout the paper.}

\subsection{Robustness of structural properties}
\label{subsec:z_BD}

H\"{a}u{\ss}ler et al (in prep.) demonstrates that the multi-band fitting used by \megamorph allows the SEDs of individual bulge and disc components of simulated galaxies to be recovered for even faint objects ($m_{r} < 20$ mag in the GAMA data). Whereas single-band fitting recovers the same SED for both the bulge and disc of a given galaxy, \megamorph's multi-band fitting shows bulges and discs to have different SEDs, even for faint galaxies.

\begin{figure*}
\centering
\includegraphics[width=0.8\textwidth]{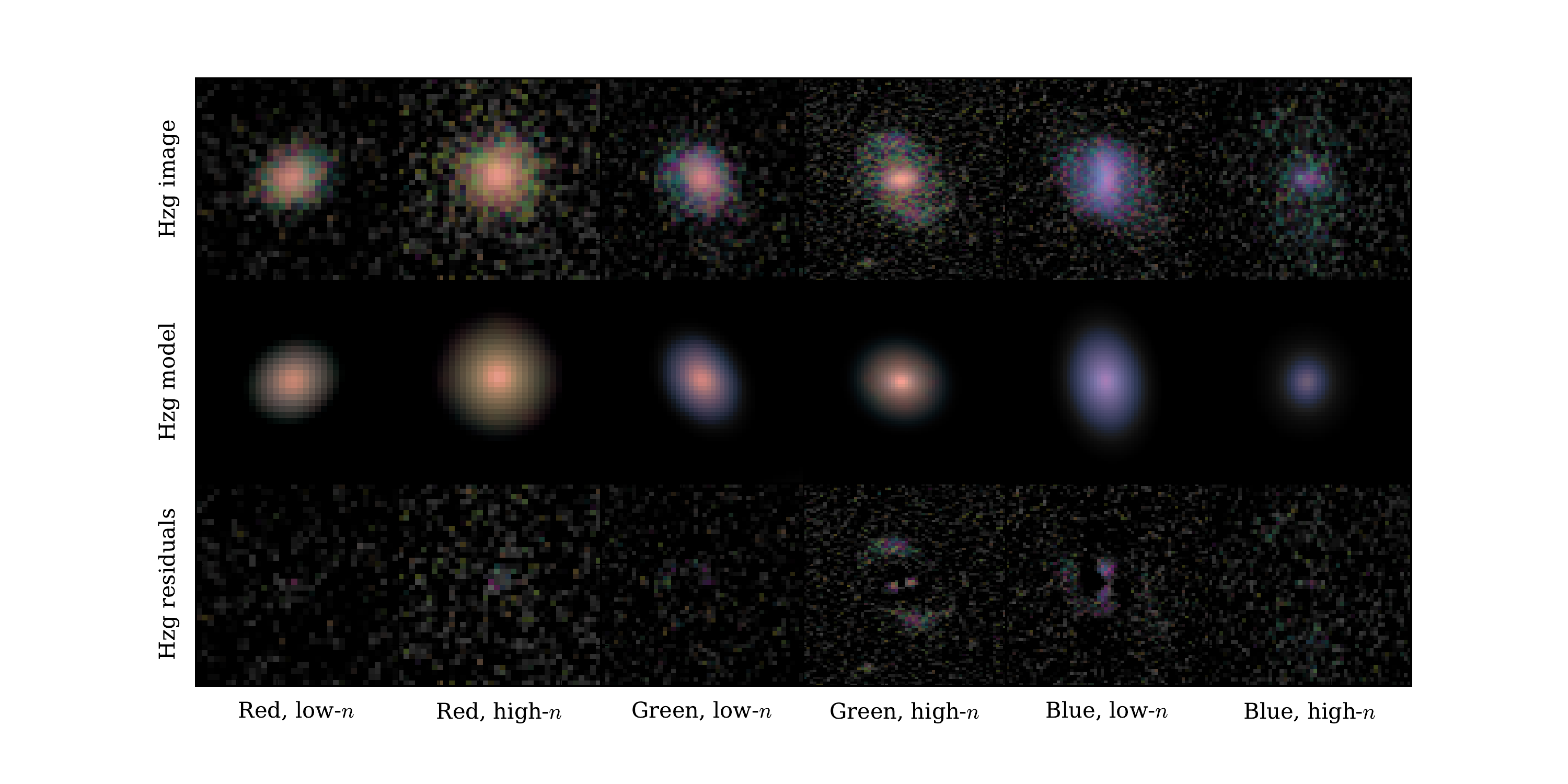}		
\caption{$Hzg$ image, single-\sersic model and residuals for an example galaxy in each of our colour/\sersic index subsamples.
\label{fig:SS_images}}
\end{figure*}

\begin{figure*}
\centering
\includegraphics[width=0.8\textwidth]{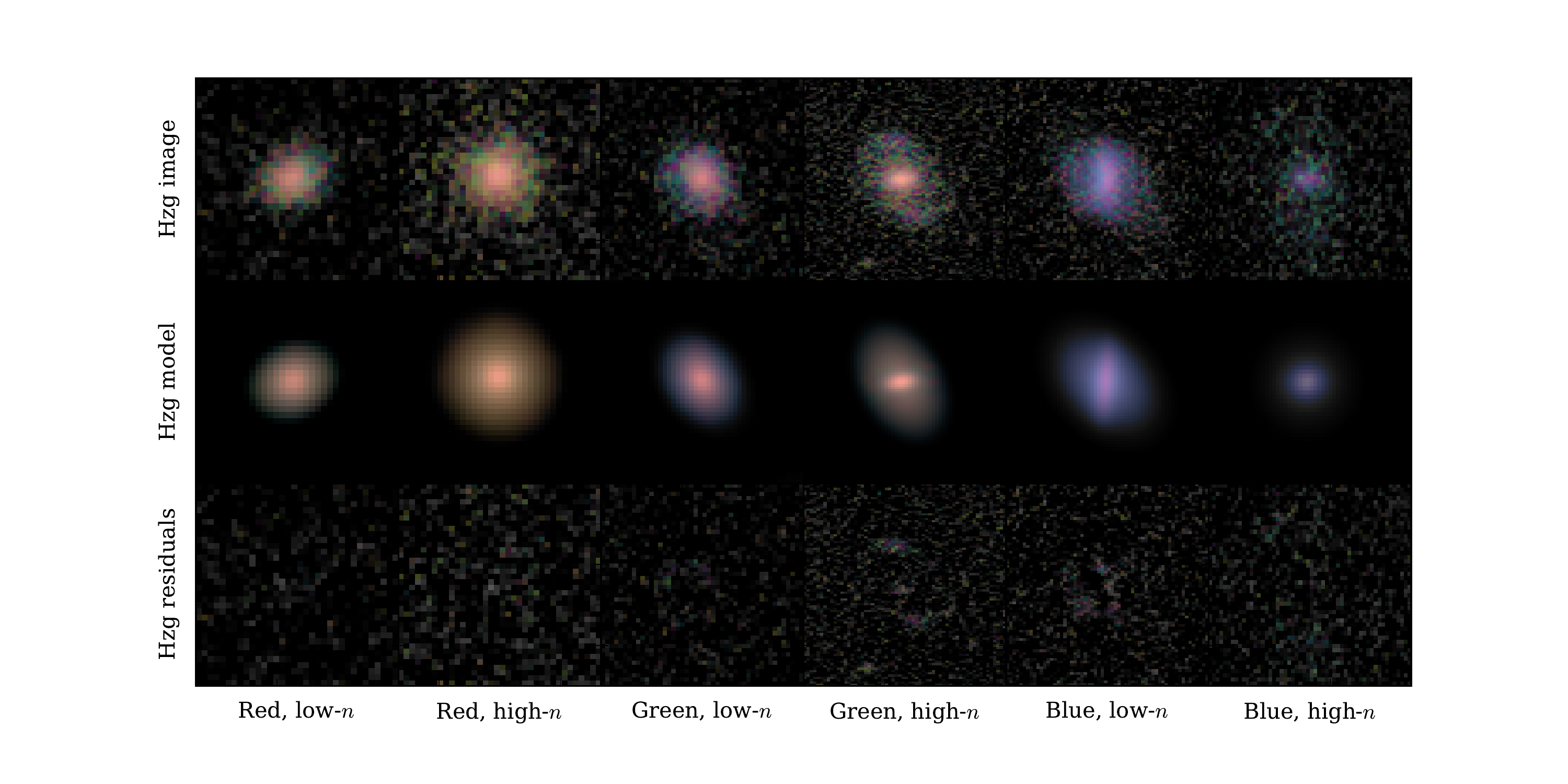}		
\caption{$Hzg$ image, bulge+disc model and residuals for an example galaxy in each of our colour/\sersic index subsamples.
\label{fig:BD_images}}
\end{figure*}

In Figures \ref{fig:SS_images} and \ref{fig:BD_images} we present three-colour ($Hzg$) images, models and residuals for \rk{example} galaxies in our six subsamples to show how well they are fit by single-\sersic and bulge-disc models respectively. By visually comparing the residuals of Fig. \ref{fig:SS_images} with Fig. \ref{fig:BD_images} we can see that the bluer objects (particularly the `$green$' high-$n$ and `$blue$' low-$n$) are \rk{slightly} better fit by a bulge and a disc than by a single-\sersic profile, which indicates that, as expected, these galaxies can generally be thought of as two-component objects. Our `$red$' galaxies are well fit by either a single- or two-component model, but adding a second component does improve the residuals. \rk{Although the residuals do not all visibly improve in the cases shown, overall the two component fits better represent most galaxies in all subsamples, as indicated by the consistent and contrasting sizes, \sersic indices and colours of the two components.}

In our previous work \citep{Kennedy2015} we have tested the single-\sersic measurements in our volume-limited samples for trends with redshift, which may arise due to biases with worsening resolution or signal-to-noise ratios. Although we measured small changes in \N and \R with redshift, we found that these were negligible compared to the differences between galaxy samples. Therefore, our results, including the strikingly different behaviour of high- and low-$n$ galaxies, are robust to redshift effects.
Here we similarly test the resilience of bulge and disc properties considered in this paper.

	\begin{figure}
		\centering
		\includegraphics[width=0.5\textwidth]{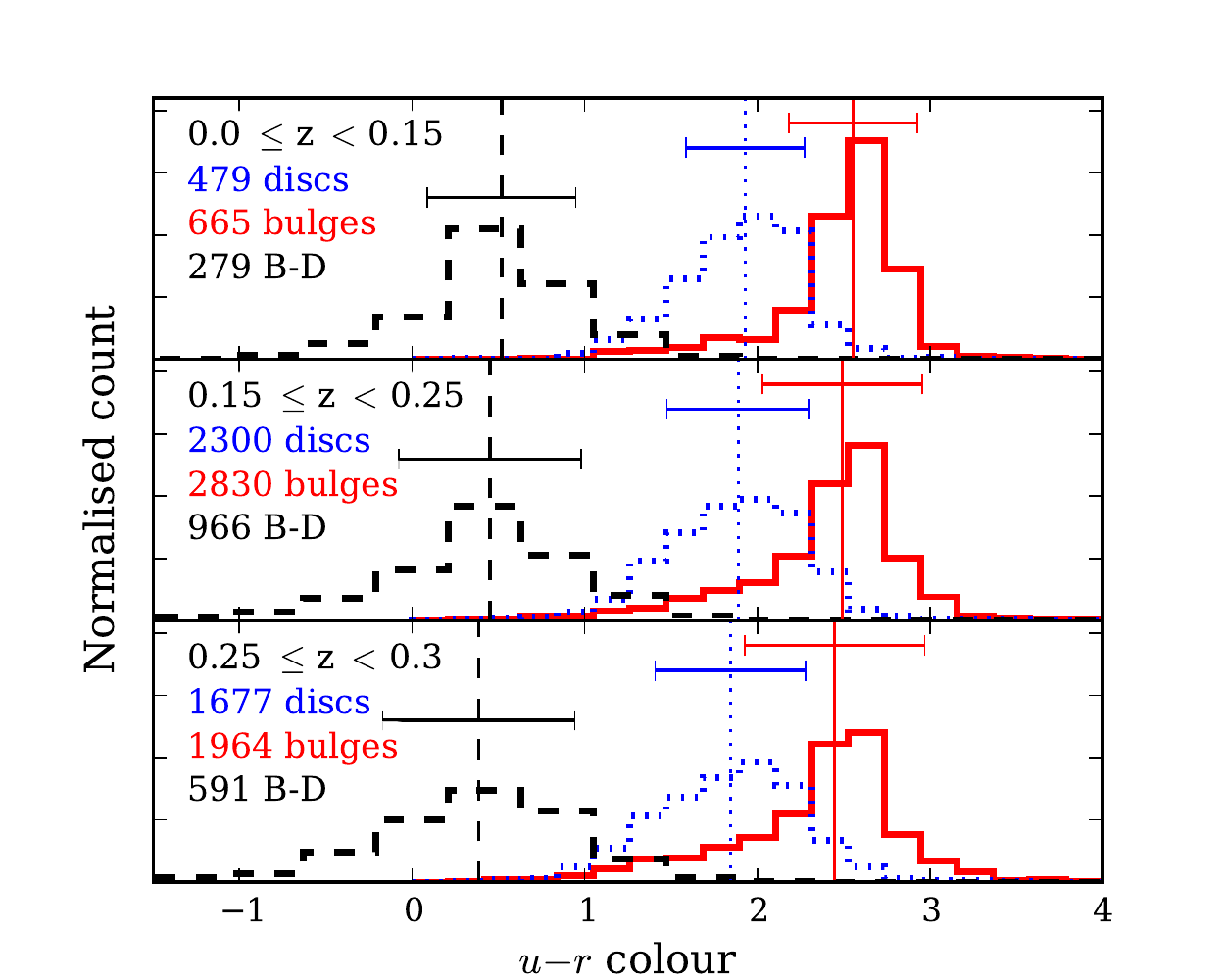}		
		\caption{The distributions of $u-r$ colour for bulges (red, solid lines) and discs (blue, dotted lines) in the cleaned V14 catalogue.  Black dashed lines show $B-D$ colours, \rk{i.e. the difference between the $u-r$ colour of the bulge and the $u-r$ colour of the disc in a given galaxy.} Each panel is restricted to $M_{r} < -21.2$ and different redshift ranges, as labelled. Median $u-r$ colours for each distribution are indicated by vertical lines, with standard deviations marked as error bars. Overall, we see that the difference between bulge and disc colours remains constant regardless of redshift.
        \label{fig:BD_Z_hist}}
	\end{figure}

Figure \ref{fig:BD_Z_hist} demonstrates the redshift dependence of bulge and disc $u-r$ colours. The bulge and disc distributions are distinct in all redshift bins, with bulges typically found to be redder than discs by 0.65 mag. At lower redshifts the colours of bulges and discs become very slightly redder by $\sim 0.1$ mag, likely due to aging stellar populations and declining star-formation rates over this 2 Gyr timescale.  Kolmogorov--Smirnov (KS) tests indicate a significant difference between the colour distribution of both bulges and discs between redshift samples, but these differences can be considered small compared with the width of the distributions. To determine whether an offset between redshift bins can be considered `small' we sum the standard deviations of the widest and narrowest distributions in quadrature. We then find the difference in the median value of $u-r$ colour in the highest and lowest redshift bins, as a fraction of the summed standard deviation. The  offset is 17.5\% of the distribution widths, which can be considered small.  Furthermore, the colour separation between bulges and discs (shown by a black dashed line) is maintained, strongly supporting the consistency of our decompositions over a wide range of signal-to-noise and resolution.
This is remarkable given that the bulges are unresolved for many of our high-$z$ objects (see Fig. \ref{fig:N_Re_Bulges_Vulcani}).

\begin{figure}
		\centering
		\includegraphics[width=0.5\textwidth]{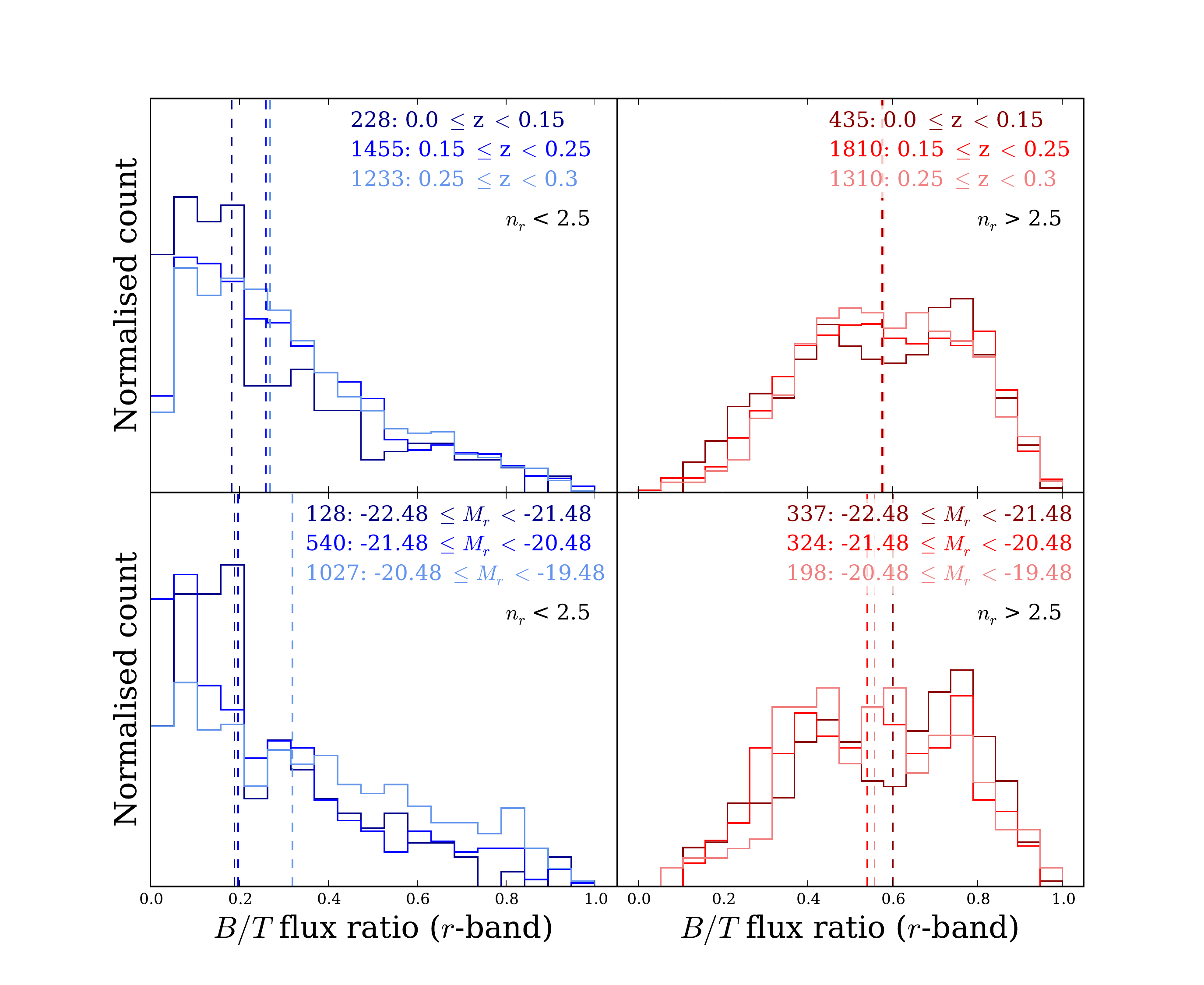}		
		
		\caption{Redshift and luminosity dependence of $B/T$ ratio for high- and low-$n$ galaxies. All galaxies which meet our basic cleaning criteria are shown here; this includes discs with very faint bulges, and vice versa, to give an accurate impression of the range of $B/T$ ratios exhibited by our sample. For the majority of galaxies the overall distribution of $B/T$ flux ratios is similar at different redshifts.  The $B/T$ flux ratio of high-$n$ galaxies show no significant dependence on luminosity, whereas the brighter the low-$n$ galaxy, the lower its $B/T$ flux ratio.
        \label{fig:BT_hist}}
		\end{figure}

We also test the dependence of $B/T$ flux ratio on redshift. Normalised histograms can be seen in the upper panels of Fig. \ref{fig:BT_hist}, showing the distribution of $B/T$ for high- and low-$n$ galaxies in three redshift bins. As seen in  \citet{Kennedy2015} for the redshift dependence of galaxy properties, the two highest redshift bins exhibit almost identical trends, whilst the lowest redshift bin shows slightly different behaviour, with more low-$n$ galaxies  exhibiting $B/T \sim 0.1$  flux ratios than the $z > 0.15$  samples.
It should also be noted that these low-$z$ bins contain far fewer galaxies than the high-$z$ bins.

\section{Results}

\begin{figure}
		\centering
		\includegraphics[width=0.5\textwidth]{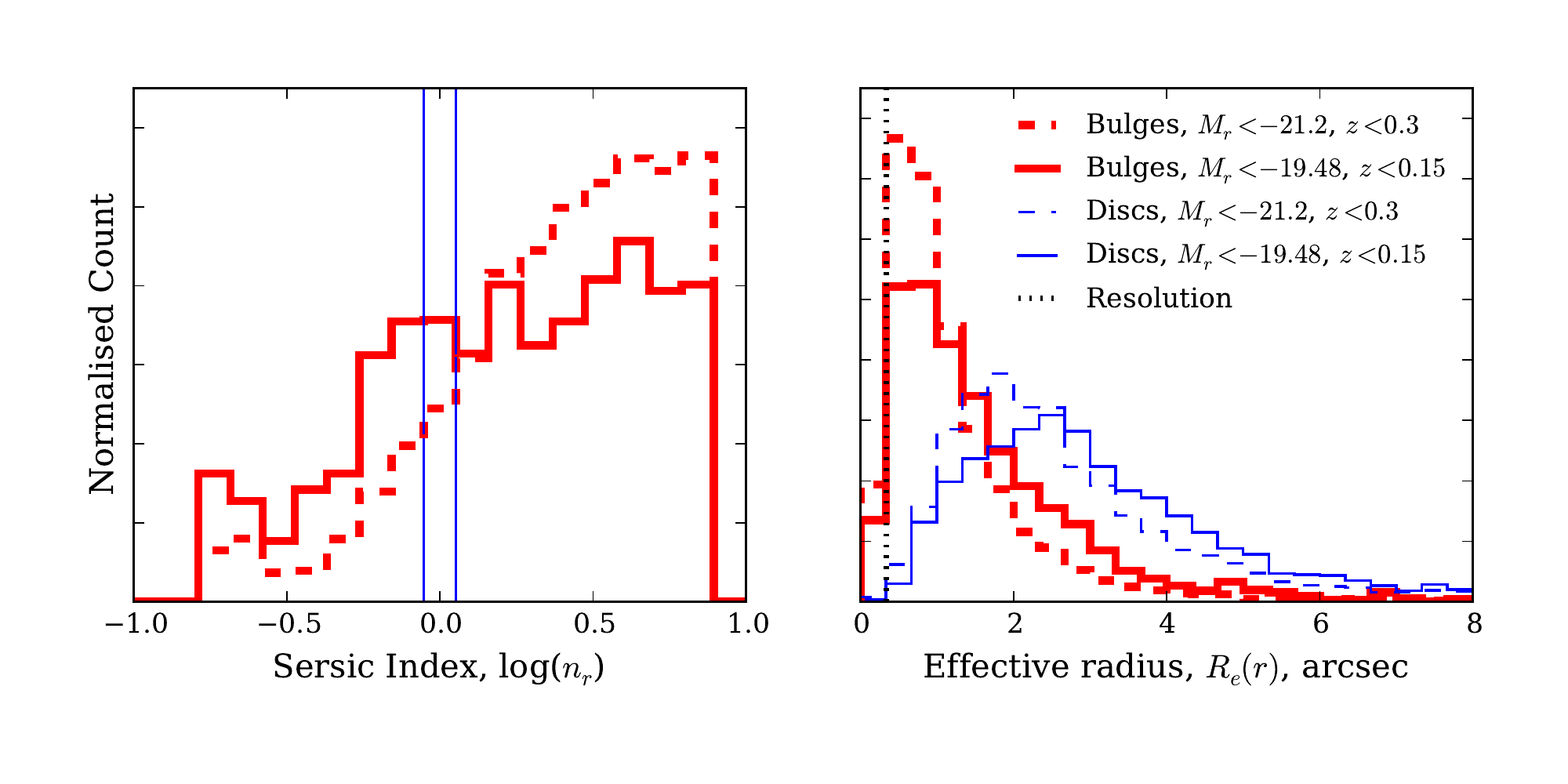}		
		
		\caption{$n_{galaxy}$ and $\re(galaxy)$ distributions for bulges (red) and discs (blue) in our `low-redshift' sample (solid lines) and the V14 sample (dashed lines). The resolution ($\re \ga 1.5$ pixels or 0.339 arcsec) is represented as a vertical dotted line in the right-hand panel. 
        \label{fig:N_Re_Bulges_Vulcani}}
		\end{figure}

To illustrate that the components from our bulge-disc decompositions generally do correspond to the usual notion of `bulges' and `discs' we show their $n$ and $\re$ distributions in Fig. \ref{fig:N_Re_Bulges_Vulcani}. By definition our discs have a \sersic index of 1, whilst bulges adopt a much wider range of \sersic indices. A large proportion (32\%) of the bulges in our low-redshift sample have lower \sersic indices than discs. This could be due to the largest, brightest galaxies being `over-fit', or galaxies with faint bulges being wrongly fitted (i.e. some disc light being attributed to the bulge).  The presence of bars could also be a factor here; we do not correct for the possible presence of bars in our sample cleaning, which could have \sersic indices as low as $\sim 0.5$ \citep{Aguerri2005,Laurikainen2007,Gadotti2011}.  \rk{Different distributions for the V14 and low-redshift samples are expected, since the two samples cover different magnitude ranges and distances. The behavior seen in this work does not change significantly depending on the sample used.}

Our bulges and discs also cover the expected relative values of effective radius; there are few small discs, but many that extend out to large radii, whilst bulges generally have smaller effective radii (in $\sim 90\%$ of cases) and none extend as far out as the largest discs.

\begin{figure}
		\centering
		\includegraphics[width=0.5\textwidth]{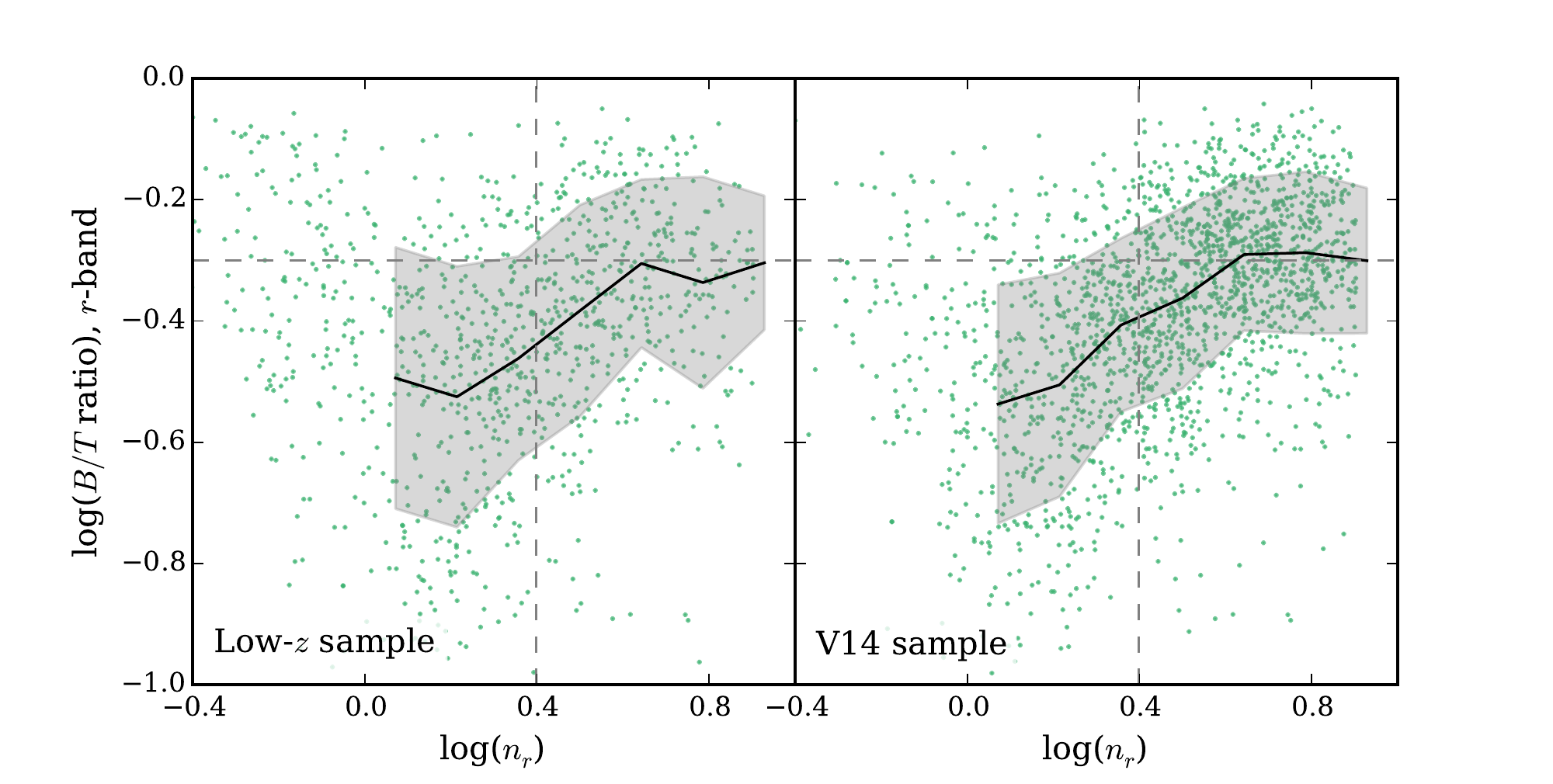}		
		
		\caption{The relationship between $n_{r}(galaxy)$ and $B/T$ for two-component galaxies in the low-$z$ sample (left-hand panel) and the V14 sample (right-hand panel). Overplotted as grey dashed lines are the low-/high-$n$ cut at $n_{r} = 2.5$, and the division between disc-/bulge-dominated galaxies at $B/T = 0.5$.  Solid black lines show the median fitted values whilst the shaded areas show the $1\sigma$ dispersion. We see a positive correlation between $B/T$ and \sersic index, implying that we can generally think of high-$n$ galaxies as being bulge-dominated, and vice versa.
        \label{fig:n_vs_BT}}
		\end{figure}

To ascertain whether our subsample of $n > 2.5$ galaxies corresponds to bulge-dominated galaxies (i.e. $B/T > 0.5$), we show in Fig. \ref{fig:n_vs_BT} the relationship between \sersic index and B/T flux ratio. Although there is a large scatter, it can be seen that there is a positive correlation between the two properties, implying that we can, to a certain extent, think of high-$n$ galaxies as generally being bulge-dominated, and vice versa.

\subsection{The wavelength dependence of single-\sersic models }
\label{subsec:wave_dep}

In this Section we compare the general structural parameters of our galaxies in order to determine whether the trends seen in V14 are supported by bulge-disc decompositions. V14 observed a change in single-\sersic index with wavelength for low-$n$ galaxies, and suggested that this may be due to the lower \sersic index of a galaxy's disc being observed in bluer wavebands, and the higher \sersic index of its bulge being observed in redder wavebands. Similarly, V14 postulates that the small change in \N seen for high-$n$ galaxies may be due to the one-component nature of these objects, while the change in \R seen for this subsample could be due to a number of different stellar populations superimposed on one another, each with a different effective radius.

\subsubsection{The wavelength dependence of \sersic index and effective radius (\N \& \R) vs. $B/T$}
\label{subsubsec:six_panel_plot}

\begin{figure*}
		\centering
		\includegraphics[width=\textwidth]{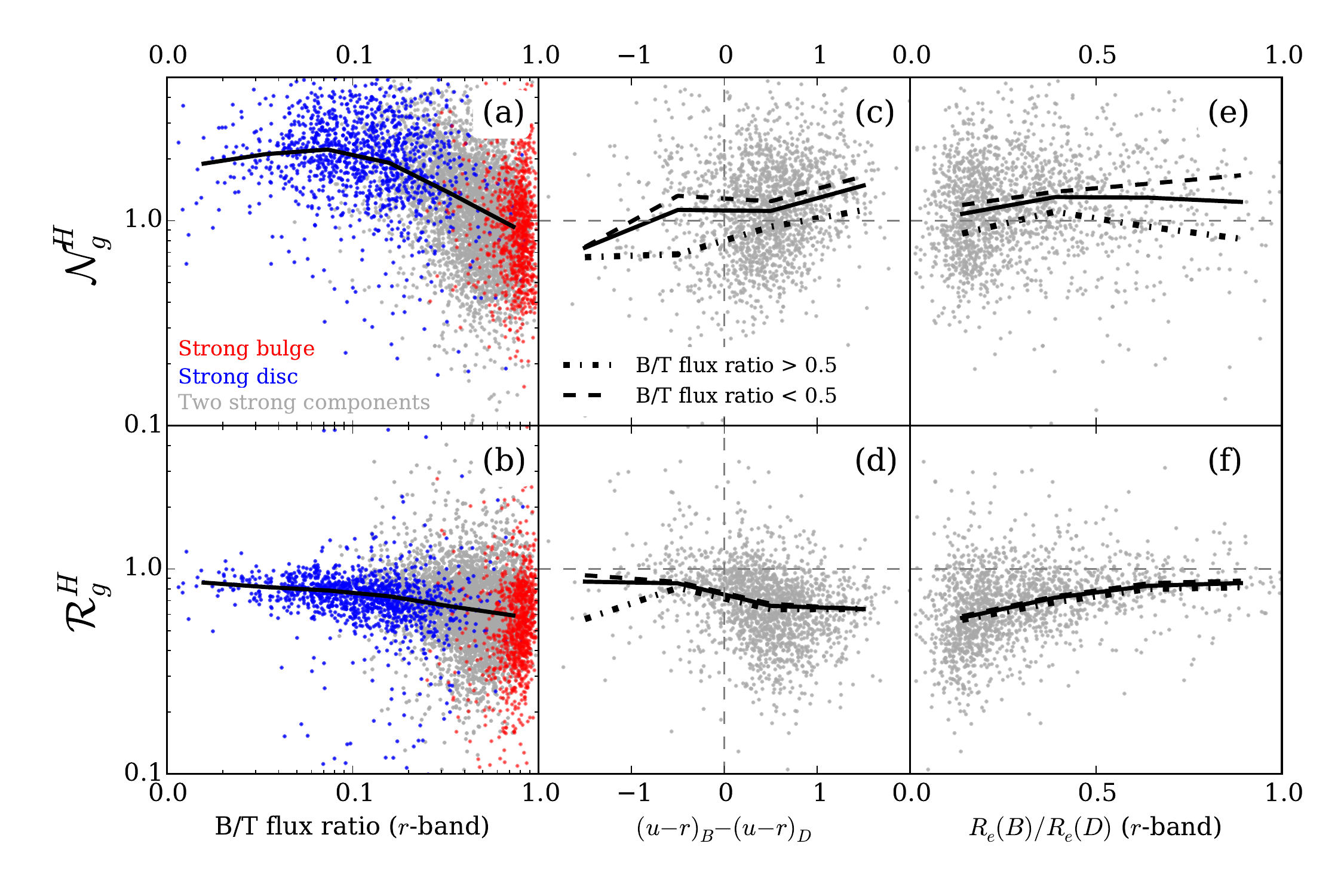}
		\caption{Upper panels show variation in \sersic index with wavelength; lower panels show variation in effective radius with wavelength, for the V14 volume-limited sample. Overlaid in black on each panel is the median \N[H][g]/\R[H][g] for the whole sample. Panels (a) \& (b) show the relationship between \N and \R vs $B/T$ flux ratio in the $r$-band; grey points show galaxies that exhibit both a bulge and disc of similar magnitudes; \rk{red points show galaxies that have only \textit{strong} bulges; blue points show galaxies that have only \textit{strong} discs (see section~\ref{subsec:Comp_selection} for more on this definition)}. See section~\ref{subsubsec:six_panel_plot} for an explanation of why there are some `\textit{strong} bulge' galaxies with low $B/T$ and vice versa. Panels (c) \& (d) show $B-D$ colour difference, whilst panels (e) \& (f) show the bulge:disc size ratio. The dashed lines in these panels show the median for galaxies with a $B/T$ flux ratio $> 0.5$, whilst the dot-dash lines show the median for galaxies with a $B/T$ flux ratio $< 0.5$. The wavelength dependence of $n$ gives a more reliable indication of a galaxy's internal structure than \R, and galaxies with similarly coloured components exhibit a weaker dependence on \R than galaxies in which the bulge is redder than its corresponding disc. We also note that there appears to be little dependence of \N on $B/D$ size ratio, however there is a stronger correlation with \R; as expected, the smaller the size ratio, the stronger the wavelength dependence of $\re$.
        \label{fig:six_plots}}		
		\end{figure*}

In Fig.~\ref{fig:six_plots} we show the relationships between \N and \R versus the relative luminosity, colour and size of the bulge and disc.  Galaxies for which we have \textit{strong} measurements of both the bulge and disc (grey points) are distinguished from those with only a \textit{strong} disc (blue points) or only a \textit{strong} bulge (red points).  See Section \ref{subsec:Comp_selection} for more details on this selection of `\textit{strong}' galaxy components. In Fig.~\ref{fig:six_plots}(a)  we show the relationship between \N and $r$-band bulge-to-total ratio, $B/T$. From the arguments in V14 we expect that galaxies with a high $B/T$ (and particularly bulge-only galaxies) will display $\mathcal{N} \sim 1$, as they are dominated by a single component, containing one population of stars. Panel (a) confirms that galaxies with $B/T \gg 0.5$ exhibit $\mathcal{N} \sim 1$, albeit with some scatter. Furthermore, V14 anticipate that galaxies with two roughly equal components, corresponding to $B/T \sim 0.5$, should have $\mathcal{N} > 1$, as a result of the higher \sersic index bulge becoming more dominant at redder wavelengths.  This is also supported by our bulge-disc decompositions. However, a deviation from the predictions of V14 comes with disc-dominated systems (with low $B/T$). Such galaxies were expected to exhibit $\mathcal{N} \sim 1$, because they are dominated by a single component.  However, on the contrary, they consistently display high values of \N. The wavelength dependence of \sersic index appears to depend not on whether a system has one or two components, but whether or not a significant disc is present.
We note that there are some galaxies with only a \textit{strong disc} which appear to have high $B/T$ flux ratios, and some galaxies with only \textit{strong} bulges which have low $B/T$ flux ratios.  \rk{Such cases arise when the second component is rejected due to the cleaning criteria in section 2.1.} These extreme cases occur because the classification of a \textit{strong} component requires a disc to be no more than three magnitudes fainter than its corresponding bulge, or vice versa. This translates to the luminosity of that disc being no less than 6\% of the luminosity of the corresponding bulge.  Hence, it is possible to have \textit{strong} bulges with $B/T = 0.06$, and \textit{strong} discs with $B/T = 0.94$ \rk{in cases where a component has been discarded due to cleaning criteria detailed in section~\ref{subsec:K_corr_etc}}.

Similarly, the relationship between \R and $B/T$ ratio can be seen in panel (b) of Fig. \ref{fig:six_plots}. Most galaxies display $\mathcal{R} < 1$, such that they appear smaller in the $H$-band than the $g$-band. Bulge-dominated systems exhibit the largest departures from unity, but also the largest scatter. This corresponds with the results for high-$n$ galaxies from V14. Galaxies with a $B/T \lesssim 0.2$ are disc-dominated and are likely to correspond to V14's low-$n$ galaxy samples. Panel (b) of Figure \ref{fig:six_plots} shows that these galaxies, as in V14, have \R closer to one than their high $B/T$ counterparts; their radii change less with wavelength.

\begin{figure}
	\centering
	\includegraphics[width=0.45\textwidth]{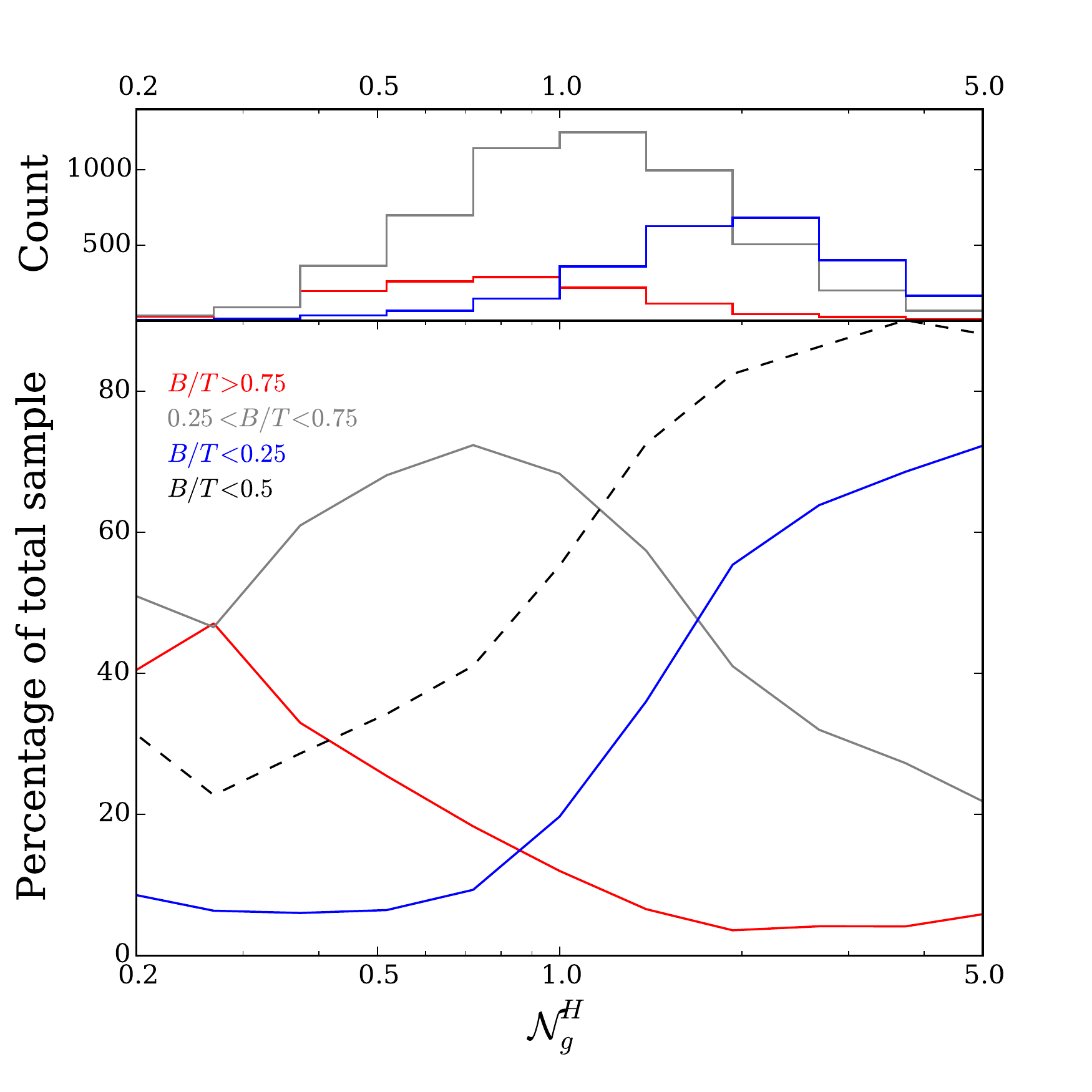}
	\caption{Upper panel shows blue, grey and red histograms giving the number of $B/T < 0.25$, $0.25 < B/T < 0.75$ and $B/T > 0.75$ galaxies respectively, for bins of \N. The lower panel shows what percentage of the whole population lies in each $B/T$ subsample at a given value of \N. We also show with a black dashed line the percentage for $B/T < 0.5$ galaxies over our range of \N. The corresponding $B/T > 0.5$ line would be a mirror image of this, and has therefore been omitted from this plot. We can see that although it would be difficult to determine whether a galaxy has a prominent bulge using \N, we can see that above \N[H][g] $\sim 1$ more than half the galaxies have prominent discs, or $B/T < 0.25$.
	\label{fig:BT_3_bins}}
	\end{figure}

From these trends we are able to estimate the likelihood of a galaxy having a bulge and/or disc at a given value of \N. In Fig.~\ref{fig:BT_3_bins} we show the percentage of galaxies at a given \N which have $B/T < 0.25$, $0.25 < B/T < 0.75$ or $B/T > 0.75$. Although galaxies with $B/T > 0.25$ can be present at all values of \N, we see that galaxies with prominent discs (i.e. $B/T < 0.25$) account for more than half the population at \N $\gtrsim 2$, while less than 10\% of the population beyond \N $\sim 2$ have $B/T > 0.75$. We have included in the lower panel of  Fig.~\ref{fig:BT_3_bins} a black dashed line showing the percentage of $B/T < 0.5$ galaxies in the sample over our range of \N. We see that at \N $\gtrsim 0.9$ ($\gtrsim 2$) we expect 50\% (80\%) of our population to be disc-dominated. We can therefore use \N to determine how likely it is that a given galaxy has a prominent disc, although selecting galaxies in combination with the \sersic index in a single band would be most effective \citep[see][]{Vika2015}.

\subsubsection{\N \& \R vs B-D colour difference and B/D size ratio}

In addition to their relative luminosity, we would expect other aspects of the bulge and disc to influence the overall wavelength dependence of galaxy structure.  If the two components have strongly contrasting colours, then the relative dominance of overall structural parameters should vary dramatically with wavelength. In cases where a galaxy's \sersic index is larger in redder wavelengths we will see $\mathcal{N} > 1$, and vice versa. Similarly, $\mathcal{R} > 1$ means a galaxy appears smaller at redder wavelengths.
Panels (c) and (d) of Fig.~\ref{fig:six_plots} shows \N and \R versus the difference between the $u-r$ colours of bulge and disc, where both components are well-constrained.  As the colour difference widens (in the typical sense of the bulge being redder than the disc), \N and \R do depart further from unity: galaxies exhibit peakier (higher-$n$) and smaller profiles at longer wavelengths.

Similarly, the relative sizes of the bulge and disc should affect the structural behaviour.  Panels (e) and (f) of Fig.~\ref{fig:six_plots} demonstrate the relationship between \N and \R versus the ratio of bulge and disc size, $\re(B)/\re(D)$, in the $r$-band. In panel (e) we see that the relative size of bulge and disc has little or no effect on \N.

There is, however, a positive correlation in panel (f); galaxies with smaller $R_{e}(B)/R_{e}(D)$ display a stronger wavelength dependence of single-\sersic effective radius.

This appears to meet our expectations: the more pronounced the difference in the size of the bulge and disc, the greater the overall decrease in size from blue to red.

There are two particularly interesting aspects of these results.  Firstly, the trends of \N versus $B-D$ colour difference and size ratio are offset for different $B/T$, while for \R they are very similar. Thus, \N is dominated by the effect of $B/T$, while \R appears to be driven by the relative size and colour of an extended `disc' component, irrespective of its relative luminosity.
 
Secondly, bulge-only systems lie at values of \R associated with the largest bulge$-$disc colour contrast.
This is consistent with trends in overall colour: galaxies with larger bulge$-$disc colour difference or larger $B/T$ tend to be redder in overall colour (see Fig.~\ref{fig:B-D_colours}).
This matches the findings of V14, in that red, high-$n$ galaxies display a dependence of size on wavelength that is stronger than bluer, more disc-like galaxies.  Drawing on the literature, V14 postulate that this behaviour is the result of accretion, preferentially to the galaxy outskirts via minor mergers, of younger or more metal-poor stars.  In the case of our bulge-disc decompositions, the blue outskirts implied by the single-\sersic fits of red, high-$n$ galaxies are either too faint to be constrained or modelled by an extended blue disc. For most cases where a disc is significantly detected, it must be associated with the usual thin disc of spiral galaxies.  However, fascinatingly, the same trends in \R and \N continue to galaxies where the disc is no longer discernable.

\subsection{$u-r$ colour distributions for bulges \& discs}
\label{subsec:B_D_colour}

\begin{figure*}
	\centering
	\includegraphics[width=\textwidth]{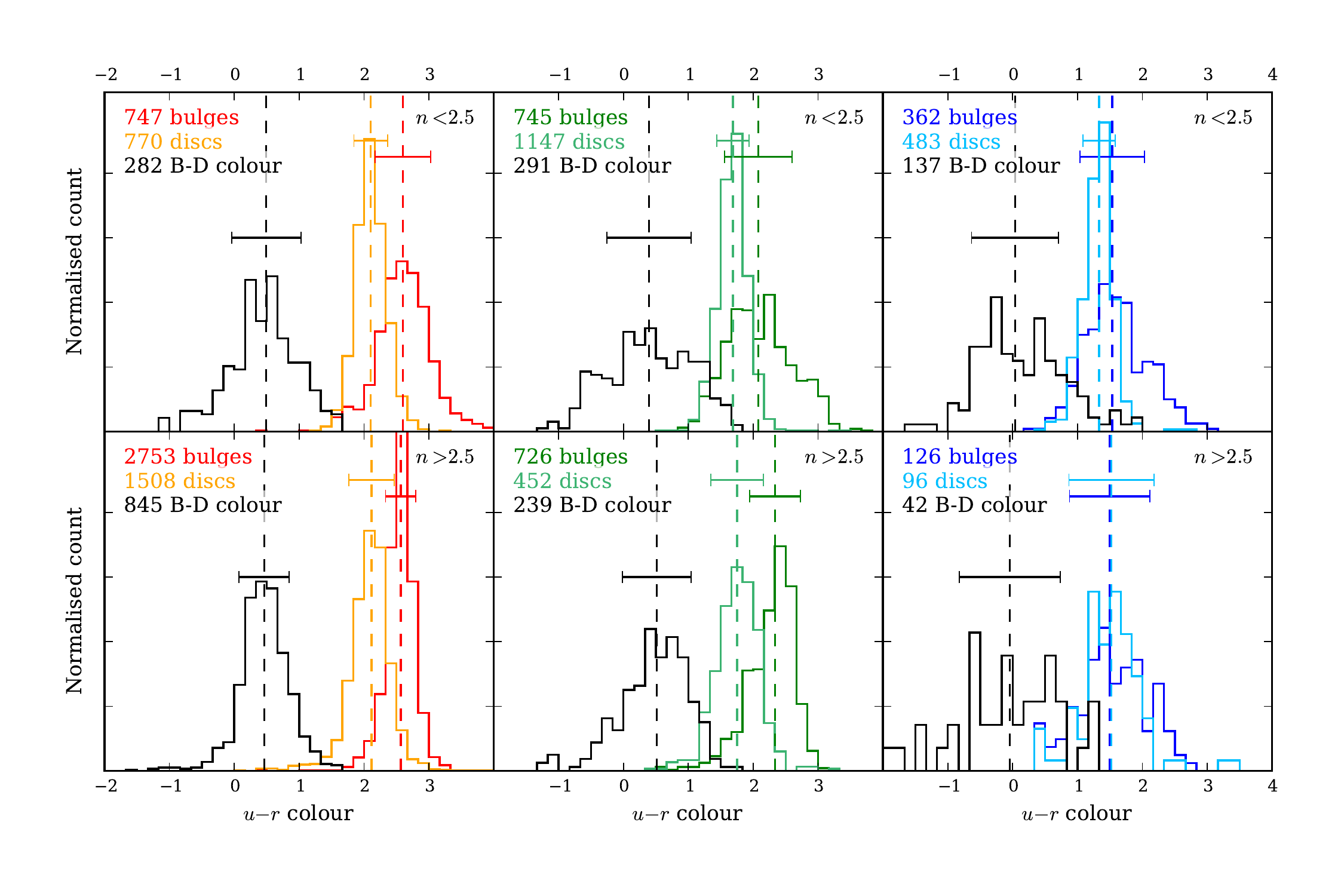}
	\caption{Normalised histogram showing the relative colour distributions of the bulges and discs of a volume-limited sample of $M_{r} < -21.2, z < 0.3$ galaxies, binned by \sersic index and overall galaxy colour, defined as follows: $u-r > 2.1$ = $red$, $1.6 < u-r < 2.1$ = $green$, $u-r < 1.6$ = $blue$. The $B-D$ colour \rk{(i.e. the difference between the $u-r$ colour of the bulge and the $u-r$ colour of the disc within a given galaxy)} is plotted in black, with negative values indicating a redder disc than bulge, and positive values indicating a redder bulge than disc. The median $B-D$ colour is plotted as a vertical black dashed line, and the standard deviation of each sample is plotted in the corresponding colour. See table \ref{table:Fig6} for exact values. Bulges are generally redder than their overall discs, and this colour difference correlates with the overall colour of the galaxy; the bluer the galaxy, the closer the colours of these two components.
	\label{fig:B-D_colours}}
	\end{figure*}

We have already seen the colour distributions for bulges and discs in Fig. \ref{fig:BD_Z_hist}.  As anticipated, we see that these two components display distinct colour distributions.
To draw more meaningful conclusions from the bulge and disc $u-r$ colours of galaxies in our sample, and to allow meaningful comparison with the single-\sersic work presented in V14, we must study the same subsamples with the added detail of our bulge-disc decompositions.
Figure \ref{fig:B-D_colours} shows the $u-r$ bulge and disc colours, and the colour difference of the two components \rk{($B-D = (u-r)_{b} - (u-r)_{d}$)}, for galaxies divided by \sersic index and colour. Median colours are overlaid as dashed lines, and are given in Table \ref{table:Fig6}. As the overall galaxy $u-r$ colour moves from red-green-blue, bulges and discs become closer in colour. For red, low \sersic index galaxies there is a narrower peak of blue discs compared to the wider distribution of redder bulges. For blue galaxies, the peaks of the bulge and disc distributions overlap, although the relative widths of the distributions are consistent with those of red and green galaxies.

\begin{table}
\centering
	\begin{tabular}{ l | c  c  c  c  c  c  c }
		& $\mu_{B}$ & $\sigma_{B}$ & $\mu_{D}$ & $\sigma_{D}$ & $\mu_{B-D}$ & $\sigma_{B-D}$ \\ \hline
		Red, low-n   			& 2.56 & 0.46 & 1.94 & 0.29 & 0.54 & 0.53 \\
		Red, high-n 			& 2.54 & 0.27 & 2.05 & 0.38 & 0.46 & 0.42 \\
		Green, low-n  			& 1.84 & 0.54 & 1.54 & 0.27 & 0.35 & 0.68 \\
        Green, high-n  			& 1.94 & 0.49 & 1.61 & 0.42 & 0.34 & 0.63 \\
        Blue, low-n  			& 1.44 & 0.54 & 1.25 & 0.30 & -0.01 & 0.72 \\
        Blue, high-n  			& 1.30 & 0.61 & 1.45 & 0.82 & -0.24 & 0.79 \\
	\end{tabular}
	\caption{The median colour of bulges ($\mu_{B}$) and discs ($\mu_{D}$) in Fig. \ref{fig:B-D_colours}, the median $B-D$ colour difference ($\mu_{B-D}$), and the standard deviations on these values ($\sigma_{B}$,$\sigma_{D}$,$\sigma_{B-D}$).
		\label{table:Fig6}}
\end{table}

The $B-D$ colours, plotted in black, show that the redder the overall galaxy, the greater the difference between the colour of the disc and the bulge. Blue galaxies show a wider distribution of $B-D$ colours, but the peak is very close to 0, showing that these blue galaxies tend to have bulges and discs with similar colours.

The bulges and discs of high \sersic index galaxies have distributions with similar widths to one another. The bulges and discs also follow the low \sersic index trend of becoming closer in colour the bluer the overall galaxy is.

The observation that bulges are consistently redder than their associated discs for the majority of our sample could imply that the formation histories of these two components are linked; rather than all bulges being intrinsically red and all discs being intrinsically blue, we see a colour difference within a given galaxy.

\begin{figure*}
	\centering
	\includegraphics[width=\textwidth]{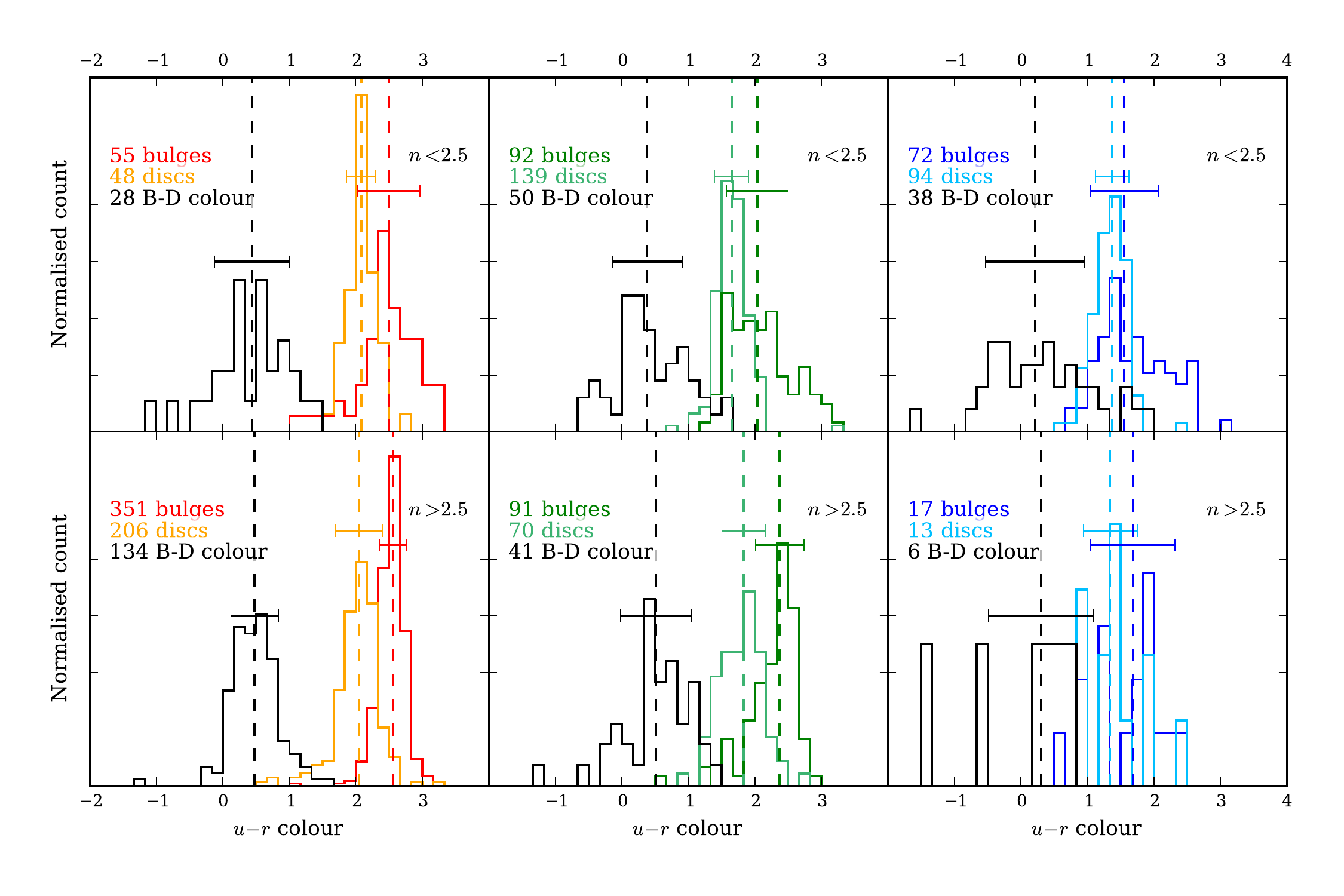}
	\caption{Face-on only ($q_{r} > 0.9$): Normalised histogram showing the relative colour distributions of the bulges and discs of a volume-limited sample of $M_{r} < -21.2, z < 0.3$ galaxies, as in Figure \ref{fig:B-D_colours}. The trends seen here are also present in Figure \ref{fig:B-D_colours}, which suggests that the inclination effects of dust are not driving these trends.
    }
	\label{fig:B-D_colours_faceon_2}
	\end{figure*}

To assess whether the trends we see here could be due to dust, we show in Figure \ref{fig:B-D_colours_faceon_2} the colour distribution of bulges and discs for face-on galaxies only.  We see that the trends observed in Figure \ref{fig:B-D_colours} are also seen in Figure \ref{fig:B-D_colours_faceon_2}, suggesting that the inclination effects of dust do not drive these trends.

\subsection{Bulge \& disc colour distributions with galaxy type}
\label{subsec:B_D_morph}

\begin{figure*}
	\centering
	\includegraphics[width=\textwidth]{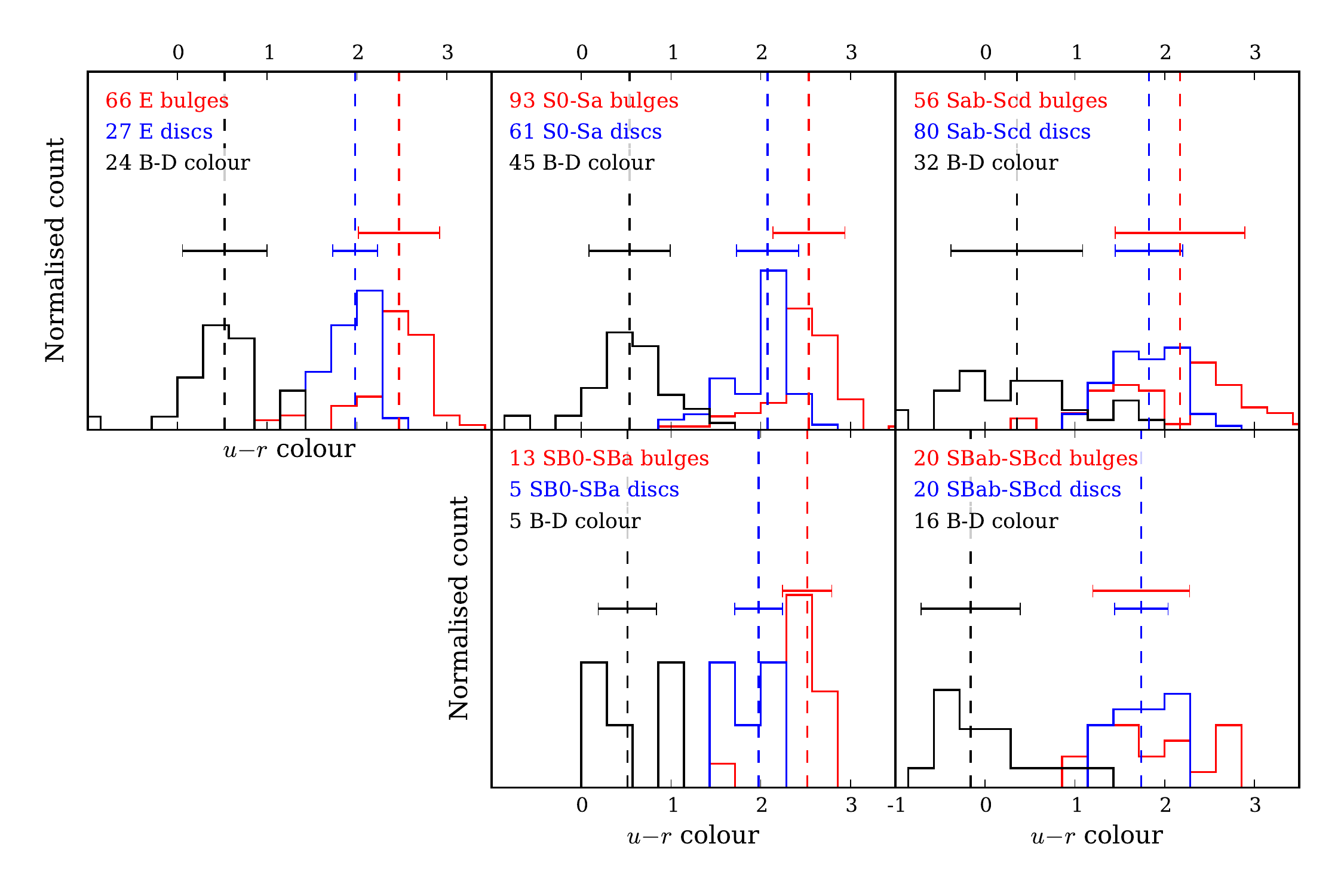}
	\caption{Normalised histogram showing the relative colour distributions of the bulges and discs of a volume-limited sample of $M_{r} < -19.48, z < 0.15$ galaxies, binned by visual morphology (visually classified by \citet{Kelvin2014}).  The $B-D$ colour is plotted in black, with negative values indicating a redder disc than bulge, and positive values indicating a redder bulge than disc. The median $B-D$ colour is plotted as a vertical black dashed line. The majority of galaxies have a bulge which is redder than its corresponding disc, irrespective of visual morphology. Even galaxies which have been visually classified as elliptical can show this $B-D$ colour difference.
	\label{fig:B-D_morph_low_z}}
	
	\end{figure*}

In the analysis of Fig. \ref{fig:B-D_colours} we made assumptions about the connection between the populations seen in these figures and the subsamples used in V14 (i.e. that red, high-$n$ galaxies correspond to our usual notion of elliptical galaxies, whilst we think of the `green' low-$n$ population as late-types). To assess how robust our assumptions are, in Fig. \ref{fig:B-D_morph_low_z} we plot the $u-r$ colour of galaxies binned according to their morphological classifications for our low-$z$ sample.  These classifications are presented in \cite{Kelvin2014}, and are a sample of 3727 galaxies with $M_{r} < -17.4$ and in the redshift range $0.025 < z < 0.06$, taken from the GAMA survey and visually classified into E, S0-Sa, SB0-SBa, Sab-Scd, SBab-SBcd, Sd-Irr and little blue spheroid classes.

Elliptical galaxies have a similar $u-r$ colour difference to red galaxies of both low- and high-n (as expected). Barred galaxies tend to show a smaller B-D colour difference than comparable non-barred galaxies. 
This trend has been studied by papers that do bulge-disc-bar decomposition \citep{Barazza2008,Weinzirl2009,Masters2010}. It most likely appears because the free \sersic function that is supposed to fit the bulge is fitting both the bulge and the bar \citep{DeGeyter2014}; consequently the stellar population of the bulge appears to be more blue that it actually is (\citealt{Peng2002,Sanchez2011}). It is also interesting to note that, when na\"\i vely fitting a bulge and a disc to all galaxies in our sample, even galaxies that have been visually classified as `elliptical' appear to contain both a \textit{strong} bulge and disc. This phenomenon has been seen before; \citet{Krajnovic2013} find kinematic evidence for discs in early-type galaxies and \citet{Huang2012} show that bright, nearby elliptical galaxies can be well-fitted with three discs of different sizes, whilst \citep{Naab2001} show that most ellipticals show evidence of a disc component containing approx. 10-20\% of the luminosity of the elliptical component.

\subsection{Luminosity dependence of bulge \& disc properties}
\label{subsec:B_D_lum}

Studying the luminosity dependence of bulge \& disc colours and flux ratios give us insight into the physical properties of our sample.
The lower panels of Fig.~\ref{fig:BT_hist} show normalised histograms of $B/T$ flux ratio for our three magnitudes bins. For $n_{r} > 2.5$ we do not see a significant difference in the $B/T$ flux ratio distribution with magnitude. In the $n_{r} < 2.5$ panel, however, we see a change in shape of the distribution with magnitude; the distribution of the brightest galaxies appears slightly `peakier' around $B/T \sim 0.1$ than for fainter galaxies.  The observation that the brighter galaxies are more disc-dominated is not necessarily surprising, as we could be seeing proportionally more star formation.

\begin{figure*}
		\centering
		\includegraphics[width=\textwidth]{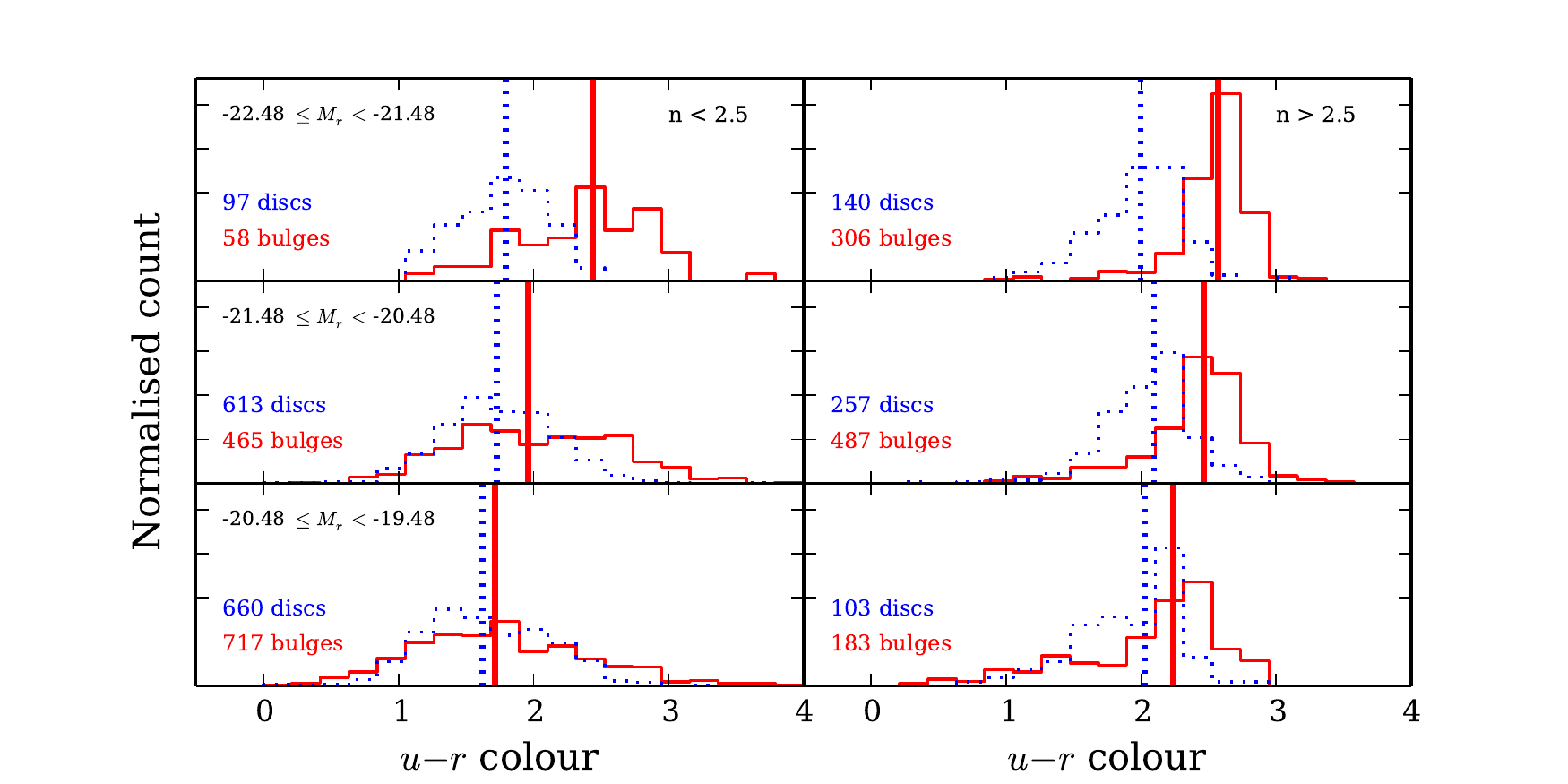}		
		\caption{Luminosity dependence of bulge \& disc colours, plotted as red solid lines and blue dotted lines respectively.  Median $u-r$ colours for each sample are overlayed as thicker vertical lines. Left- and right-hand panels show low- and high-$n$ galaxies respectively. Both the high- and low-$n$ populations show a trend of fainter galaxies appearing to be bluer in overall colour.
        \label{fig:BD_Mr_hist}}
		\end{figure*}

Figure \ref{fig:BD_Mr_hist} shows the luminosity dependence of bulge and disc $u-r$ colours for high- and low-$n$ samples. Low-$n$ systems experience a change in bulge and disc colour with luminosity; the fainter the galaxy, the closer in colour the bulge and disc appear, and the bluer the galaxy overall.  High-$n$ galaxies see a similar, although less pronounced, trend.  It is interesting to note that in all populations the colours of the discs are comparatively unchanged by luminosity, whereas the bulges get significantly bluer (by up to 0.7 mag, in the case of low-$n$ galaxies).
A certain amount of the trends seen for fainter galaxies may be influenced by the fitting process; because the galaxies are so faint, the components cannot be as easily separated and therefore some properties of the bulge may be interpreted as that of the disc and vice versa. This is explored further in H\"au{\ss}ler et al. (in prep.).

\section{Discussion}

In this paper we have presented multi-band bulge-disc decompositions for a sample of 10491 galaxies and tested that key quantities ($B/T$ and colours of bulge and disc) are robust to the effects of redshift.  These decompositions have been used to study how bulge and disc properties relate to the structural behavior measured using wavelength-dependent single-\sersic models.  We have then focused on how the relative colours of bulges and discs vary with overall galaxy properties.

\subsection{Comparison of observed trends of \R and \N with other studies}

The wavelength dependence of single-\sersic structure was measured by V14, and summarised in terms of the fractional variation in \sersic index and effective radius between the $g$- and $H$-bands, which we denote \N[H][g] and \R[H][g].

To recap, a value of unity for \R and \N means that a galaxy appears to be the same size, and has the same \sersic index, between $g$- and $H$- wavebands.  If $\mathcal{R} < 1$, as in the majority of galaxies, that object will appear larger in bluer wavebands.  Conversely, in the more unlikely case in which $\mathcal{R} > 1$, a galaxy would appear bluer in the centre and redder at larger radii.  The variation in \sersic index with effective radius, \N, is greater than unity in cases where a galaxy's \sersic index is larger (more `peaky') in redder wavelengths, and smaller (`flatter') in bluer wavelengths.  The reverse is true for $\mathcal{N} < 1$ galaxies.

We find that the different \N and \R distributions of high- and low-$n$ galaxies can be attributed to specific trends in the relative luminosity, colour and size of their constituent bulges and discs.  The high-$n$ galaxy population are generally more bulge-dominated, and have \N closer to unity.  However, while these galaxies are often considered uniformly-red, single-component systems, we find they display \R substantially below unity: their effective radii are much smaller at longer wavelengths.  Our decompositions attribute this to the presence of an extended, bluer, component, at least reasonably described by an exponential profile.  \cite{Dullo2013} and \cite{Graham2015} have revealed that many local, massive galaxies are in fact lenticular galaxies with large 2D discs rather than spheroids with large 3D envelopes.  In some cases the presence of a disc, including spiral features, may be visually confirmed.  In the remaining cases, the extended component may be a faint disc or more spherically distributed material, although the trends in the properties of this component, as well as \N and \R, with $B/T$ suggest some continuity in its origin. \rk{The properties of these discs are consistent with what is seen in studies of kinematics \citep[e.g.][and references therein]{Emsellem2011,Krajnovic2011}}

Our low-$n$ galaxy population is dominated by discs, i.e. low $B/T$, which we have found to be consistently associated with $\mathcal{N} > 1$, even in the case of apparently disc-only systems.  This population also displays \R somewhat closer to unity, as a result of less pronounced differences in the colours and sizes of their bulge and disc components.

The luminosity dependence of \N versus \R in \citet{Kennedy2015} can be understood as primarily due to lower-luminosity galaxies (at a given $B/T$ or $n$) having closer bulge and disc colours, and hence \N and \R closer to unity.  The variation in \N and \R with overall colour in V14 mainly appears to result from the correlation between colour and luminosity; the more luminous galaxies tend to have a greater difference between their bulge and disc colours, which results in a greater change in structural properties with wavelength.

\subsection{Comparison of observed trends of component colours with other studies}

\rk{We remind the reader that in this study we have applied a bulge-disc decomposition to all galaxies in our sample, regardless of whether there is a physical need for two components. We have done this primarily to avoid building a dichotomy into our results, but this has also resulted in some interesting observations, in particular the sample of visually classified ellipticals we see in Fig.~\ref{fig:B-D_morph_low_z} which have \textit{strong} discs associated with them.  We do, however, apply a cleaning algorithm to distinguish between potentially unnecessary components and eliminate bulges which are significantly fainter than their corresponding disc (and vice versa). See section~\ref{subsec:Comp_selection} for more details. The purpose of our cleaning is to avoid considering the properties of components that make an insignificant contribution to the galaxy light.} 
Our analysis has shown that the redder and more luminous a galaxy, the greater the difference between the colour of its bulge and disc.
\cite{Hudson2010} have performed bulge-disc decompositions simultaneously in $B$ and $R$ bands for $\sim 900$ galaxies in nearby clusters, and find that the reddest (and brightest) galaxies have a larger gap between bulge and disc colours. Although we find a small dependence of disc colour on magnitude in our low-$n$ population, this effect is minimal compared to the strong dependence of bulge colour on magnitude, which is also consistent with the findings of \cite{Hudson2010}. In agreement with this, \cite{Head2014} also observe a greater difference between bulge and disc colours for brighter objects in their sample of $S0$ galaxies.

Regardless of visual morphology we see that bulges are consistently redder than their associated discs; \cite{Lackner2012} (and references therein) find that discs around classical bulges are redder than lone discs or discs around pseudo-bulges, which supports our observation that bulge and disc colour are correlated.

The work of \cite{Peletier1996}, however, suggests that the colour variations from galaxy to galaxy are much larger than the colour differences observed between the bulges and discs of individual galaxies, for a sample of inclined, bright, early-type spirals.  This is somewhat at odds with our work, which suggests that the overall colour of a galaxy is driven by the relative colours of the bulge and disc.  Nonetheless, \cite{Peletier1996} find a $B-D$ colour, $\Delta(U-R)$, of $0.126 \pm 0.165$, which (within error) is consistent with both our study and \cite{Cameron2009}.

By looking at the colours of bulges and discs, we can infer their star formation histories and eventual quenching.  The negative colour gradients seen in the majority of galaxies (e.g. \citealt{Prochaska2011, Roediger2011} and references therein) tells us that older, redder stars tend to lie in the central, collapsed regions of galaxies, whilst the (rotationally supported) outskirts of a galaxy are generally dominated by younger, bluer stellar populations.  On average over our six subsamples, bulges are 0.285 mag redder than their corresponding discs, and are indeed smaller and more concentrated.  With our detailed analysis, however, we are able to see that this mean magnitude difference is a combination of the larger and smaller B-D colour differences seen in $red$ and $blue$ populations respectively.

\section{Summary}
\rk{We remind the reader that in this work we fit a bulge and disc ($n = free$ and $n = 1$ respectively) to all galaxies in our sample. We make no attempt to remove objects for which a 2-component fit is inappropriate, nor do we substitute single-\sersic measurements in these cases. We do, however, remove bulges which are more than 3 magnitudes fainter than their corresponding discs, and vice versa (see section~\ref{subsec:Comp_selection} for more details). We also note that we use the terms `bulge' to refer generally refer to the central component of a galaxy (thus, bars, lenses, pseudo bulges, classical bulges, and their superpositions, are all swept up in this term), whilst we use `disc' to refer to a more extended component with an exponential light profile.}

\begin{itemize}
	\item The difference between bulge and disc colours of both high- and low-$n$ galaxies remains constant regardless of redshift (see Fig. \ref{fig:BD_Z_hist}). The overall distribution of $B/T$ flux ratios is similar at different redshifts, with perhaps a slightly higher proportion of low-$z$ galaxies appearing to be disc-dominated in the lowest redshift bin.    
    \item \N \& \R (single-\sersic wavelength dependence) give us information about a galaxy's bulge and disc properties (see Figs. \ref{fig:six_plots} and \ref{fig:BT_3_bins}):
    \begin{itemize}
    	\item The wavelength dependence of \sersic index, \N, indicates whether an object is likely to contain a disc; \N \textgreater 1 = likely to have a disc present, \N \textless 1 = bulge-dominated galaxy
    	\item The wavelength dependence of $\re$ is a less effective classifier of structure than \N. Little change in $\re$ with wavelength suggests that a disc is present, whereas more change in $\re$ with wavelength suggests that the galaxy could be bulge-dominated.
        \item A strong wavelength dependence of $n$ is correlated with a redder B-D colour, i.e. a larger difference between the colour of the bulge and the colour of the disc.
        \item Irrespective of the $B/T$ flux ratio of the system, galaxies with similarly coloured components exhibit a weaker dependence of $\re$ on wavelength than galaxies with a bulge redder than its disc.
        \item For the entire V14 sample we see little dependence of \N on $B/D$ size ratio. However, once we split our sample into bulge-dominated and disc-dominated galaxies ($B/T$ flux ratio \textgreater 0.5 and $B/T$ flux ratio \textless 0.5, respectively), we see that the disc-dominated galaxies show an increase in \N with $B/D$ size ratio, whilst the bulge-dominated population decreases in \N with increasing $B/D$ size ratio.
        \item The relative size of the bulge and disc have little effect on \N, but there is a correlation with \R; as one would expect, galaxies with a smaller $R_{e}(B)/R_{e}(D)$ display a stronger wavelength dependence of single-\sersic effective radius.
    \end{itemize}
    
    \item Bulges are generally redder than their associated discs (see Fig. \ref{fig:B-D_colours}), regardless of the overall galaxy colour or \sersic index. The bulge and disc are closer in colour for galaxies that are bluer in overall colour. Rather than all bulges being red and all discs being blue, there appears to be a colour difference within a given galaxy. For example, the median colour of \textit{green} high-$n$ bulges is actually bluer than the median colour of \textit{red} high-$n$ discs, which is what we might expect if bulges in bluer galaxies are likely to be pseudo-bulges.
    \item Regardless of morphology, the majority of galaxies exhibit a bulge that is redder than its corresponding disc (see Fig. \ref{fig:B-D_morph_low_z}). This is particularly interesting in the case of galaxies that have been visually classified as ellipticals, yet still appear to have a \textit{strong}, comparatively blue, disc component.
    \item For the low-$n$ population, brighter galaxies exhibit a lower $B/T$ flux ratio, whereas the high-$n$ population shows no significant change in $B/T$ with luminosity (see Fig. \ref{fig:BD_Mr_hist}).  Bulges and discs get closer in colour for fainter galaxies (regardless of $n$). For both high- and low-$n$ populations, the fainter the galaxy, the bluer its overall colour.

\end{itemize}

\section{Acknowledgements}
This paper is based on work made possible by NPRP award 08-643-1-112 from the Qatar National Research Fund (a member of The Qatar Foundation).

GAMA is a joint European-Australasian project based around a spectroscopic campaign using the Anglo-Australian Telescope. The GAMA input catalogue is based on data taken from the Sloan Digital Sky Survey and the UKIRT Infrared Deep Sky Survey. Complementary imaging of the GAMA regions is being obtained by a number of independent survey programs including GALEX MIS, VST KiDS, VISTA VIKING, WISE, Herschel-ATLAS, GMRT and ASKAP providing UV to radio coverage. GAMA is funded by the STFC (UK), the ARC (Australia), the AAO, and participating institutions. The GAMA website is http://www.gama-survey.org/.
R.K. acknowledges support from the Science and Technology Facilities Council (STFC).
SPB gratefully acknowledges the receipt of an STFC Advanced Fellowship.
BV acknowledges the financial support from the World Premier International Research Center Initiative (WPI), MEXT, Japan and  the Kakenhi Grant-in-Aid for Young Scientists (B)(26870140) from the Japan Society for the Promotion of Science (JSPS)
SB acknowledges the funding support from the Australian Research Council through a Future Fellowship (FT140101166)

We thank the anonymous referee for their constructive comments.

\bibliographystyle{mn2e}
\bibliography{4Dec14}

\end{document}